\documentclass[sigconf]{acmart}
\usepackage{multirow}
\usepackage{lineno}
\usepackage{todonotes}
\usepackage{colortbl}
\usepackage{caption}
\usepackage{subcaption}
\usepackage{booktabs}

\usepackage{xcolor}
\usepackage{soul}
\usepackage{xr}

\makeatletter
\newcommand*{\addFileDependency}[1]{
  \typeout{(#1)}
  \@addtofilelist{#1}
  \IfFileExists{#1}{}{\typeout{No file #1.}}
}
\makeatother

\newcommand*{\myexternaldocument}[1]{%
    \externaldocument{#1}%
    \addFileDependency{#1.tex}%
    \addFileDependency{#1.aux}%
}
\myexternaldocument{samples/appendix}

\AtBeginDocument{%
  \providecommand\BibTeX{{%
    \normalfont B\kern-0.5em{\scshape i\kern-0.25em b}\kern-0.8em\TeX}}}

\setcopyright{acmcopyright}
\copyrightyear{2018}
\acmYear{2018}
\acmDOI{XXXXXXX.XXXXXXX}

\usepackage{xspace}
\newcommand{\framework}{ACTOR\xspace}

\copyrightyear{2023}
\acmYear{2023}
\setcopyright{acmlicensed}\acmConference[IUI '23]{28th International Conference on Intelligent User Interfaces}{March 27--31, 2023}{Sydney, NSW, Australia}
\acmBooktitle{28th International Conference on Intelligent User Interfaces (IUI '23), March 27--31, 2023, Sydney, NSW, Australia}
\acmPrice{15.00}
\acmDOI{10.1145/3581641.3584045}
\acmISBN{979-8-4007-0106-1/23/03}

\begin{document}

\title[The Importance of Multimodal Emotion Conditioning and Affect Consistency for ECAs]{The Importance of Multimodal Emotion Conditioning and Affect Consistency for Embodied Conversational Agents}



\author{Che-Jui Chang}
\affiliation{%
  \institution{Rutgers University}
  \state{New Jersey}
  \country{USA}
}
\email{chejui.chang@rutgers.edu}

\author{Samuel S. Sohn}
\affiliation{%
  \institution{Rutgers University}
  \state{New Jersey}
  \country{USA}
}
\email{samuel.sohn@rutgers.edu}

\author{Sen Zhang}
\affiliation{%
  \institution{Rutgers University}
  \state{New Jersey}
  \country{USA}
}
\email{sen.z@rutgers.edu}

\author{Rajath Jayashankar}
\affiliation{%
  \institution{Rutgers University}
  \state{New Jersey}
  \country{USA}
}
\email{rajath.jay@rutgers.edu}

\author{Muhammad Usman}
\affiliation{%
  \institution{King Fahd University of Petroleum \& Minerals}
  \state{Dhahran}
  \country{Saudi Arabia}
}
\email{muhammad.usman@kfupm.edu.sa}

\author{Mubbasir Kapadia}
\affiliation{%
  \institution{Rutgers University}
  \state{New Jersey}
  \country{USA}
}
\email{mubbasir.kapadia@rutgers.edu}



\begin{abstract}
Previous studies regarding the perception of emotions for embodied virtual agents have shown the effectiveness of using virtual characters in conveying emotions through interactions with humans.
However, creating an autonomous embodied conversational agent with expressive behaviors presents two major challenges. The first challenge is the difficulty of synthesizing the conversational behaviors for each modality that are as expressive as real human behaviors. The second challenge is that the affects are modeled independently, which makes it difficult to generate multimodal responses with consistent emotions across all modalities. In this work, we propose a conceptual framework, {\framework} (\textbf{A}ffect-\textbf{C}onsistent mul\textbf{T}imodal behavi\textbf{OR} generation), that aims to increase the perception of affects by generating multimodal behaviors conditioned on a consistent driving affect.
We have conducted a user study with 199 participants to assess how the average person judges the affects perceived from multimodal behaviors that are consistent and inconsistent with respect to a driving affect.
The result shows that among all model conditions, our affect-consistent framework receives the highest Likert scores for the perception of driving affects.
Our statistical analysis suggests that making a modality affect-inconsistent significantly decreases the perception of driving affects. We also observe that multimodal behaviors conditioned on consistent affects are more expressive compared to behaviors with inconsistent affects.
Therefore, we conclude that multimodal emotion conditioning and affect consistency are vital to enhancing the perception of affects for embodied conversational agents.
\end{abstract}



\begin{teaserfigure}
  \includegraphics[width=\textwidth]{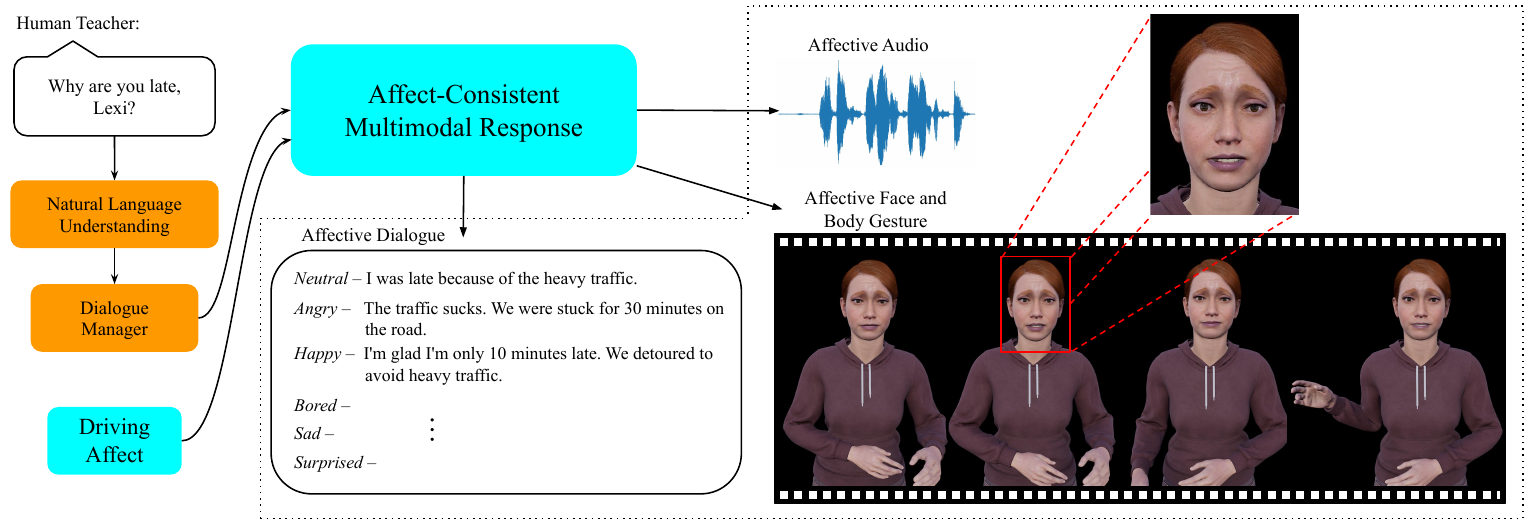}
  \caption{We propose, \framework, a conceptual framework with affect-consistent multimodal behaviors for autonomous embodied conversational agents that enhances human's perception of affects. Our framework consists of four affective modalities, dialogue (left, below center), voice (right, above center), face, and body gesture (lower right), generated from the same driving affect (lower left).}
  \label{fig:teaser}
\end{teaserfigure}

\maketitle




\section{Introduction} \label{sec:intro}
Embodied virtual agents have garnered increasing attention in the field of graphics and virtual reality as they facilitate the building of realistic, immersive environments and low-risk virtual training platforms \cite{delamarre2019aimer, badiee2015design}. Several studies \cite{mcdonnell2008evaluating, clavel2009combining, ennis2013emotion} have shown the efficacy of emotion contagion with the use of expressive virtual agents during multimodal interactions with humans. However, it remains a challenge to generate affective multimodal responses for autonomous embodied conversational agents (ECAs) that are as expressive as humans.
As it stands, creating an embodied conversational agent with expressive multimodal responses requires the effort of putting together all the modalities and synchronizing the multimodal behaviors, but the fact that the emotions are modeled separately could be another obstacle for building such an expressive framework.
In this work, we propose a conceptual framework, \framework (\textbf{A}ffect-\textbf{C}onsistent mul\textbf{T}imodal behavi\textbf{OR} generation), with multimodal emotion conditioning and affect consistency that addresses the aforementioned issues and increases human's perception of affects.

The concept of emotion conditioning refers to the capability of taking an emotion as a conditional input and generating the behavior for a modality that matches the condition. For example, an emotion-conditioned face modality may use the audio and conditioned emotion as inputs to generate an affective facial animation. On the contrary, the non-conditioned modalities, which are included in most existing embodied conversational frameworks \cite{sohn2018emotionally, nagy2021framework, dipaola2019multi}, only accept one input mode and output another mode without consideration of emotion. The notion of affect consistency refers to an integrated ECA framework being able to generate multimodal emotional behaviors with the same affect. Terminologically, we refer to the conditioned emotion for each modality as the driving affect in the \framework framework, as it is used to drive the affective behaviors.

Our affect-consistent framework consists of four modalities: dialogue, voice, face, and body. The conversational behaviors for each modality are generated given the same driving affect, as illustrated in Figure \ref{fig:teaser}.
Practically, the behaviors are generated using stylistic parameters that have been linked to each driving affect, as described in Section \ref{sec:implementation}. We describe the preliminaries of our main user study, including the design of the experiments, creation of stimulus, and the comparison models, in Section \ref{sec:userstudy}. We show the confusion matrix of the perception scores for each comparison model to evaluate the efficacy of multimodal emotion conditioning and affect consistency on affect perception and conduct ANOVA tests to report the statistical significance in affect perception under 4 model conditions, 6 driving affects, and 6 perceived affects, in Section \ref{sec:results}. We summarize the experimental results and discuss our key findings in Section \ref{sec:discussion}.


This paper makes the following contributions. 
First, we propose a conceptual multimodal framework, \framework, that resolves the two aforementioned challenges for building autonomous ECAs, including the difficulty of generating the expressive behaviors for the conveyance of emotions and the issue of stitching together the modalities in which the affects are modeled separately. Such a framework can be applied to create intelligent virtual agents in video games or deployed in human-computer interaction interfaces to increase the immersive experiences for virtual training and assistance. 
Second, we discover from our user study that affect consistency maximizes the perception of the driving affect and that an inconsistent affect in just one modality can decrease the same affect perception. 
In fact, when a modality is not emotion conditioned, the behaviors tend to be less expressive, which then dilutes the perception of the driving affect.
These findings evidence the importance of emotion conditioning and affect consistency for each modality.
Finally, an additional statistical analysis reveals several more nuanced findings. (1) The voice and face modalities contribute to affect perception more than the body modality, as the removal of a consistent affect in the body modality does not necessarily disrupt the perception of the same affect. (2) The correlations in perception scores between the affects can be explained using their valence and arousal values. These findings are important for building an ECA framework with multimodal emotional responses in which the affects and modalities play a vital role.
The video demonstration for the affect-consistent multimodal behaviors can be found in our supplementary materials.

\begin{figure*}[!t]
  \centering
  \includegraphics[width=\linewidth]{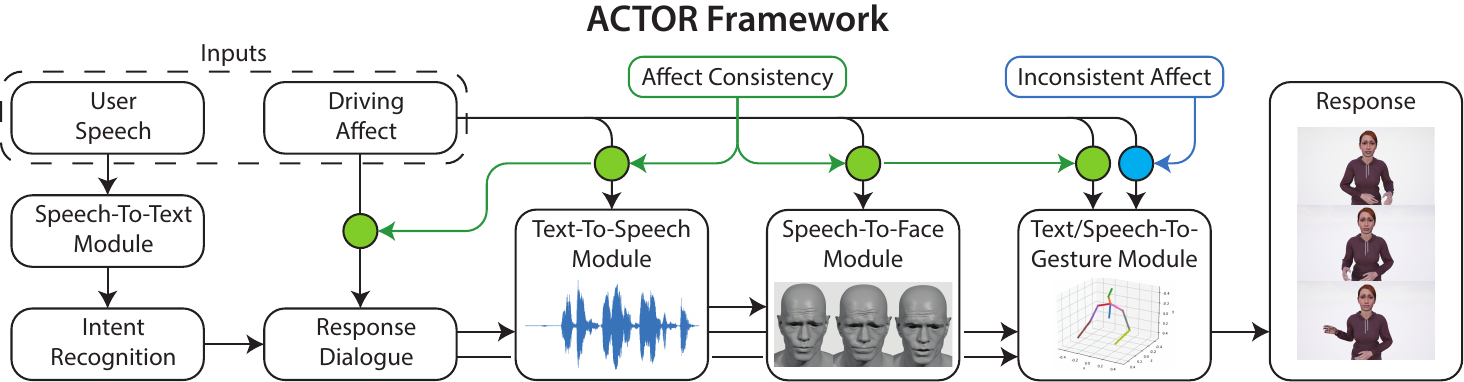}
   \caption{Our conceptual framework, \framework, with multimodal emotion conditioning and affect consistency. First the intent is recognized from the user's speech input and the corresponding sentence response is used to generate the voice for the virtual agent. The synthesized audio is then used to generate the face animation. Last, the body gestures are generated using the text and audio as inputs. The same driving affect is conditioned for all modalities to achieve affect consistency. To evaluate the influence of emotion conditioning and affect consistency on the perception of affects in the user study, we build affect inconsistent models for comparisons, where a single modality has an inconsistent affect.}
   \label{fig:actor}
\end{figure*}

\section{Related Work}
Our proposed framework and user study for affect perception are related to the coordination of the agent's perception and responses, the modeling of synchronous modalities, and the evaluation of affect-driven or personality driven agents. 
In this section, we review the literature from the following perspectives: Multimodal Conversational Agents and Affect-Driven Avatars.

\subsection{Multimodal Conversational Agents}
\subsubsection{Multimodal Communication}
Multimodal communication is a natural form between human interlocutors where audio, facial expressions, eye gazes, head movements, hand gestures, and body gestures are used to provide vivid conversational behaviors.
Likewise, the same communication strategy has been proven to be effective in increasing the realism of embodied virtual agents according to previous studies \cite{ferstl2021human, potdevin2018virtual}. Social skills including behavior matching \cite{louwerse2012behavior}, style matching \cite{hoegen2019end}, and emotion-awareness \cite{sohn2018emotionally} have been investigated and developed in virtual conversational agents. In addition, research \cite{blomsma2020intrapersonal} has shown that synchronization of the modalities, specifically speech and gestures, for multimodal ECAs is the key to improving realism and human likeness. 
We leverage the findings from these prior works by applying data-driven approaches for multimodal behaviors and synthesizing the behaviors with consistent affects.

\subsubsection{Coordination of Agent's Perception and Response}
Authoring an interactive ECA requires an integrated framework that coordinates the agent's perception and conversational responses. The interactive behavior tree was introduced in the literature \cite{shoulson2013adapt, kopp2006towards, kapadia2015evaluating, janghorbani2019domain} for such a framework. PICA \cite{falk2018pica} models the active and reactive agent behaviors for a conversational agent, with the ability to initiate conversations, parse user intents and handles responses. Typically, a Natural Language Understanding (NLU) module is the core of the dialogue management that maps user's semantic inputs to the predefined intent. According to the intent and dialogue state, a pre-authored multimodal response is played. For instance, \cite{lee2006nonverbal} generates dialogue responses by a rule-based NLU module and incorporates gestures as an additional modality. \cite{sonlu2021conversational} creates and modifies the personality-driven multimodal behaviors offline and combines the modalities to render a response to an intent. The work from \citep{sohn2018emotionally} takes multimodal perception inputs, processes the user emotion and dialogue intent, and generates the pre-authored animation response with emotion-awareness. 
Our work follows the same pipeline for intent parsing, but applies emotion-conditioned methods for multimodal response generation. 


\subsubsection{Synchronization of Modalities} \label{sec:2.1.4}
As the modalities are generated through separate channels, synchronization becomes a key issue to human perception of realism, especially between speech and gestures \cite{blomsma2020intrapersonal}. 
\cite{ferstl2020understanding} investigates the speech-gesture correlation and aims to understand the predictability of gesture from speech. 
Their follow-up work \cite{ferstl2021expressgesture} leverages the speech-gesture relationship for gesture synchronization by finding the best gesture in the dataset through feature matching. Another branch of speech-gesture synchronization studies \cite{bhattacharya2021text2gestures, habibie2021learning, ginosar2019learning, alexanderson2020style, yoon2020speech, chang2022ivi} focuses on using a parameterized network to learn mappings between the two modalities. 
For example, Speech2Gesture \cite{ginosar2019learning} detects the hand gestures from video and builds a convolutional network for gesture generation. \cite{alexanderson2020style} uses a probabilistic flow-based model for likelihood maximization for co-speech gesture generation. 
On the other hand, several research groups \cite{randhavane2019eva, bhattacharya2021text2gestures, bhattacharya2021speech2affectivegestures, yoon2020speech} have included semantic features to infer gestures. \cite{yoon2020speech} encodes audio, text, and speaker identity for the generation of gestures.
\cite{bhattacharya2021speech2affectivegestures} includes an additional seed pose as input for autoregressive affective gesture generation. For more works regarding speech-driven gesture, we direct readers to these review papers \cite{liu2021speech, wolfert2022review}.

\subsection{Affect-Driven Avatars}
\subsubsection{Affective Avatars}
The accurate modeling of affects increases the believability of a conversational agent. Prior work \cite{shvo2019interdependent} builds a computational model for affects with the psychologically-guided interplay of affective components including personality, motivation, emotion, and mood. \cite{liu2020building} conducts a user study to create psychologically-plausible facial expressions.
The increasing modeling capability of affects enables us to create ECAs capable of showing empathy and expressing emotional behaviors in terms of dialogues \cite{casas2021enhancing}, faces \cite{chen2020dynamic, liu2020building}, and gestures \cite{bhattacharya2021speech2affectivegestures}.
Previous studies include emotionally-aware avatars with affect matching \cite{sohn2018emotionally} and affect regulation \cite{yalccin2020empathy}, personality-based emotional characters \cite{sajjadi2019personality}, and an affective virtual student \cite{delamarre2019aimer} to support teacher training. 

Our framework does not incorporate affective perception but focuses on the synthesis of multimodal affect-consistent behaviors with our defined emotions.

\begin{table*}[!t]
  \caption{The mapping of the affects and the stylistic parameters in the voice \cite{ibm2015}, face \cite{omniverse2021}, and body \cite{bhattacharya2021speech2affectivegestures} modality. For example, neutral audio would increase the speaking rate by 40\%. For the face modality, the source shot represents the style of the facial expression. The smoothing parameter determines the smoothness of the face animation can be along the time axis. The strength parameter determines the degree of exaggeration. For the body modality, the seed pose ID refers to the pre-pose used for training in \cite{bhattacharya2021speech2affectivegestures}. It determines the style of the predicted pose. We randomly select one of the listed IDs in our framework.}
  \label{tab:emotive_features}
  \resizebox{\linewidth}{!}{%
  \begin{tabular}{cc|cccccl}
    \toprule
     \textbf{Modality} & \textbf{Affective Parameter} & \textbf{Neutral} & \textbf{Angry} & \textbf{Happy} & \textbf{Bored} & \textbf{Sad} & \textbf{Surprised} \\
    \midrule
    \multirow{5}{*}{Voice} 
                   & Pitch & - & 20\% & 100\% & -20\% & 20\% & 80\% \\
                   & Pitch Range     & - & 100\% & 100\% & -20 & 20\% & 40\% \\
                   & Rate            & 40\% & 50\% & 10\% & -30\% & -10\% & 20\% \\
                   & Breathiness     & - & - & -100\% & - & 100\% & -30\% \\
                   & Glottal Tension & - & 80\% & 60\% & - & -20\% & 50\% \\
    \cmidrule{2-8}
    \multirow{3}{*}{Face}
                  & Source Shot           & \texttt{p3\_neutral} & \texttt{g1a} & \texttt{g2b} & \texttt{g8b} & \texttt{g8c} & \texttt{g4b} \\
                  & Upper, Lower Smoothing     & 0.011, 0.004 & 0.012, 0.000 & 0.012, 0.005 & 0.025, 0.004 & 0.056, 0.009 & 0.014, 0.002 \\
                  & Upper, Lower Strength     & 0.600, 1.218 & 0.798, 1.608 & 1.874, 1.632 & 0.600, 1.400 & 1.594, 1.126 & 1.696, 1.330 \\

    \cmidrule{2-8}
    Body & Seed Pose IDs & \texttt{18}, \texttt{30} & \texttt{5}, \texttt{10} & \texttt{8}, \texttt{27} & \texttt{23}, \texttt{130} & \texttt{123}, \texttt{182} & \texttt{142}, \texttt{181} \\

  \bottomrule
\end{tabular}
}
\end{table*}

\subsubsection{Personality-Driven Avatars}
Building an ECA with personality-driven behaviors has become a major focus in recent years, with evidence and findings from prior research \cite{shvo2019interdependent, ishii2020impact, durupinar2016perform, castillo2018personality} to support the benefit of personality modeling. Typically, the OCEAN personality \cite{mccrae1992introduction, durupinar2016perform, castillo2018personality} is used in the literature for analysis and generation of character movements. \cite{sonlu2021conversational} does personality-specific adjustments for all conversational modalities. The voices and facial expressions for OCEAN personalities are built on top of an affect layer. Their body motions are modified according to Laban Movement Analysis.
Though our work focuses primarily on affect-driven behaviors, we apply similar adjustments to the modalities for emotion conditioning.

\section{Conceptual Framework} \label{sec:conceptual framework}
In general, the process of generating multimodal responses involves the following sequential steps: (1) dialogue generation, (2) text-to-speech, (3) audio-driven (or text-driven) facial animation, and (4) audio-driven body animation. However, such an ECA framework that generates the behaviors without emotion conditioning could lead to \textit{emotion dilution} --a multimodal response that is supposed to be emotional and expressive, but turns out to be perceived as more neutral. 
Specifically, the reason comes from the fact that each step of the sequential executions is trained separately on different domains and then concatenated in a framework. 
For example, a non-conditioning system, with a text-to-speech module followed by an audio-driven facial animation module, may have the two modules trained on different datasets. The quality of the face modality can be undermined when synthetic audio generated from a different domain is used as its input during inference. Additionally, the degree of emotional expressiveness can often be diluted in each sequential step due to the sequential execution of modules without emotion conditioning. For example, the same non-conditioning system can dilute the expressiveness in the face, even though an affective text input is given initially. 
This is because all of these data-driven modules have to decode the affect from the input and encode it into the output behavior. The learning of affects is not guaranteed, so the affect in the output modality can become diluted.
Therefore, all modalities in our proposed framework are emotion-conditioned.

It is widely accepted that the perception of affects is the most heightened when all modalities share the same affect. Prior research \cite{clavel2009combining, ennis2013emotion} regarding the effect of different modalities of a virtual character on the perception of emotions always assumed that combined modalities shared the same emotion in their study design. However, achieving affect consistency for ECAs is challenging because all modalities can model affects in different ways. The affect that is conditioned in one modality might not be used as the condition in another. Moreover, some modules do not have interpretable variables for emotion conditioning. For example, the work from \cite{karras2017audio} is able to generate expressive face animations from audio input, but the variables used to control the output expressions are not interpretable. 
Therefore, we propose the conceptual framework, \framework, with all modalities conditioned on a consistent affect, as illustrated in Figure \ref{fig:actor}.
Like other ECA frameworks, our framework follows a similar sequential execution but uses the same driving affect as the condition input for generating multimodal behaviors. We detail the implementation of our conceptual framework in the following subsections. To show the importance of emotion conditioning and affect consistency in an ECA framework, we also conduct a user study (Sec. \ref{sec:userstudy}) to compare our affect consistent model condition with the affect inconsistent models with one modality having a different driving affect.


\begin{table*}[!t]
  \caption{An example of our designed dialogue with intent, sample question, and responses for all affects.}
  \label{tab:dialogue_example}
  \begin{tabular}{cp{10.5cm}}
    \toprule
    \textbf{Intent} & Cover-up 1 \\
    \cmidrule{2-2}
    \textbf{Question} & ``Then why are you the only one in class who's late today?'' \\ 
    \midrule
    \multirow{1}{*}{\textbf{Neutral}} & \multirow{1}{*}{``I went to a doctor earlier and then came to school.''} \\
    \cmidrule{2-2}
    \multirow{1}{*}{\textbf{Angry}} & ``Well, I had to see a doctor first. That was \textit{not} my idea. My mom asked me to.'' \\
    \cmidrule{2-2}
    \multirow{1}{*}{\textbf{Happy}} & ``Well, I came here after a doctor's appointment. My headache was relieved.'' \\
    \cmidrule{2-2}
    \multirow{1}{*}{\textbf{Bored}} & \multirow{1}{*}{``I had to see a doctor first. Then come here.''} \\
    \cmidrule{2-2}
    \multirow{1}{*}{\textbf{Sad}} & ``I had to go to see a doctor. Please don't blame me for that. I'm really sorry.'' \\
    \cmidrule{2-2}
    \multirow{1}{*}{\textbf{Surprised}} & \multirow{1}{*}{``Oh! Um, my mom asked me to see a doctor first.''} \\
  \bottomrule
\end{tabular}
\end{table*}

\subsection{Implementation} \label{sec:implementation} 
We realize the conceptual framework based on existing data-driven methods and create a virtual student, Lexi, based on the Metahuman character \cite{metahuman2021}. We use pre-scripted affective text responses and apply text-to-speech synthesis, audio-to-face generation, and gesture generation from text and audio in our implementation. 
In order to create a shared affect space for multimodal emotion conditioning, we utilize a mapping from the emotions to the stylistic parameters for each modality. 
The parameters are either not annotated (e.g. latent codes learned in an unsupervised fashion) or interpretable but not directly related to emotions. A summary of the mappings is shown in Table \ref{tab:emotive_features}.
The modifications of the voice parameters are based on psychological studies \cite{schroder2001emotional} on features for emotional voices while the adjustments of the latent parameters for faces and gestures are validated in our preliminary user study (Sec. \ref{sec:pre_study}).
Based on the framework, we design two scenarios: the late scenario and the homework submission scenario. The perception of the character in our implementation is done by a speech-to-text engine followed by an intent extraction. Depending on the intent of the user, the scenarios guide users through different dialogue states and ultimately the dialogues are directed to different endings. We refer readers to our appendices for more details regarding the framework implementation and our scenario flowcharts.

The following 6 affects are used to condition the modalities: \textit{neutrality, anger, happiness, boredom, sadness, and surprise}.
We do not use all emotions from the universal model \cite{ekman1999basic} because contemptuous, disgusted, and fearful voices are not as distinguishable as others for most widely used Text-To-Speech engines, such as IBM TTS \cite{ibm2015}.

\subsubsection{Dialogue Modality}

Research from prior works \cite{Zhou2018EmotionalCM, colombo2019affect-driven} is intended for generating dialogue responses with the driving affect as the condition. However, we could not adopt these prior works for our affective dialogue generation because the generated dialogues do not fit in our two scenarios and guide users to specific branches and endings.
To this end, we designed our own dialogues. Table \ref{tab:dialogue_example} shows an example of the dialogue in the late scenario. For each dialogue state, we created 6 response sentences, which correspond to the 6 driving affects. These affective responses were then used as inputs for the generation of behaviors in other modalities.

\subsubsection{Voice Modality}

Given an input text and a driving affect, \framework synthesizes the speech for the voice modality. 
Although there exist emotion-controllable text-to-speech synthesis methods \cite{li2021controllable}, these previous studies are not applicable because of either their poor quality, domain differences, or language disparity \cite{li2021controllable, chang2020transfer}.
Instead, we use the IBM Watson Text-To-Speech Engine \cite{ibm2015} for our voice modality. The provided controllable features are \textit{pitch, pitch range, rate, breathiness, and glottal tension}. \textit{Pitch} refers to the frequency of the voice. \textit{Pitch range} specifies the variation of pitch during speech. \textit{Rate} refers to the talking speed. \textit{Breathiness} determines the amount of air produced in the sound. \textit{Glottal tension} decides how hard the voice is. A voice with higher breathiness and lower glottal tension sounds calm and soft.
We carefully set the feature values for all affects according to the review paper \cite{schroder2001emotional}. The mapping is shown in Table \ref{tab:emotive_features}.
For detailed information about the features, we refer readers to these works \cite{ibm2015, sonlu2021conversational}.

\subsubsection{Facial Animation Modality}
The face modality takes the synthesized audio and the driving affect as inputs, and then outputs a facial animation with synchronized lip movements and a matching affect. The previous work proposed by \cite{karras2017audio} builds an end-to-end model for the task using unsupervised latent codes weakly associated with affects. Recently, the work from \cite{chang2022disentangle} disentangles audio content and emotion and entangles the content with the driving affect for expressive emotion-conditioned animation synthesis. Although these works synthesize high-quality emotional facial animations, their facial motions cannot be retargeted to the Metahuman character's parametric blendshape \cite{metahuman2021}. As a result, we leverage Omniverse Audio2Face \cite{omniverse2021}, where the parametric face model is driven by input audio and stylistic parameters, including the source shot, smoothing, and strength. The source shot controls the style of the facial expression. The smoothing parameter determines the smoothness of the face animation can be along the time axis. The strength parameter determines the degree of exaggeration.
We create the mappings and link the affects with the parameters for emotion conditioning, as shown in Table \ref{tab:emotive_features}. 
The generated animations are then retargeted to the Metahuman character.

\subsubsection{Body Gesture Modality}
As discussed in Section \ref{sec:2.1.4}, body gestures can be synthesized from audio and text inputs by an end-to-end network.
For instance, Text2Gestures \cite{bhattacharya2021text2gestures} is designed to generate affective, natural-looking gestures from textual semantics, while Speech2AffectiveGestures \cite{bhattacharya2021speech2affectivegestures} leverages multimodal inputs, including text, audio, speaker identity, and a seed pose, to synthesize the affective gestures. 
Notably, the affective constraint in the latter work enables the output gestures to share the same affect with the seed pose. The seed pose is the pose sequence in the first few frames of the training segment. During inference, it can be used to generate gestures for different talking styles.
Therefore, we build our gesture modality on top of Speech2AffectiveGestures, where the dialogue and synthetic speech are passed to the pretrained model for body gesture generation.
The Speech2AffectiveGestures model was trained on TED Gesture Dataset \cite{yoonICRA19}, where only the upper bodies (10 joints) were used. The prediction of 10 joint positions are converted to the rotational angles relative to their parent joints and then retargeted to their corresponding joints in our character. 
We select the seed pose ID as the affective parameter for the body modality. The selection of the affective parameter is shown in Table~\ref{tab:emotive_features}.


\subsection{Preliminary User Study} \label{sec:pre_study}
\begin{table}[!h]
  \caption{Summary of the pairwise comparison for affects. Each row represents whether an affective behavior can be identified when paired with all other affective behaviors. 
  The \textit{average match} is defined as the percentage of \textit{match} cases plus half \textit{equal} cases.}
  \label{tab:modality_acc}
  \begin{tabular}{cc|ccccl}
    \toprule
     & Affect & Match & Equal & Mismatch & Avg. Match \\
    \midrule
    \multirow{6}{*}{Face} & Neutral & 0.55 & 0.35 & 0.10 & 0.73 \\
     & Angry & 0.64 & 0.29 & 0.07 & \textbf{0.79} \\
     & Happy & 0.82 & 0.08 & 0.10 & \textbf{0.86} \\
     & Bored & 0.63 & 0.27 & 0.10 & \textbf{0.77} \\
     & Sad & 0.54 & 0.34 & 0.11 & 0.71 \\
     & Surprised & 0.59 & 0.31 & 0.10 & 0.75 \\
    \midrule
    \multirow{6}{*}{Body} & Neutral & 0.52 & 0.38 & 0.10 & 0.71 \\
     & Angry & 0.55 & 0.43 & 0.02 & \textbf{0.77} \\
     & Happy & 0.53 & 0.35 & 0.12 & 0.71 \\
     & Bored & 0.73 & 0.18 & 0.08 & \textbf{0.83} \\
     & Sad & 0.52 & 0.25 & 0.23 & 0.64 \\
     & Surprised & 0.40 & 0.47 & 0.13 & 0.63 \\
  \bottomrule
\end{tabular}
\end{table}

We conducted a preliminary user study for the validation of our choice of affective parameters for both face and gesture modalities.
We adopted the pairwise comparison method in the study where a random pair of affective behaviors were presented at the same time. The participants were then asked which better matches the description. The two affects of the selected behaviors were queried. For example, a pair of angry and happy facial animations was presented and two questions, which animation is happy and which is angry, were asked.
The users could select either left, right, or equal as their response. The survey includes two parts, one for facial animations and the other for body movements. 
For each affect, we rendered 3 stimuli with each lasting roughly 3 seconds long. The behaviors only contained one modality and the rendered videos were without audio. The selection of the pairs and the order of the queried affects were randomized. 
We recruited 18 participants from the university, and each participant completed 30 survey questions: 15 videos for the face modality and 15 for the body modality.

We report the percentages of match, mismatch, and equal for the pairwise comparison result for the affects. Each affect is compared with all other affects and the summary of the result is shown in Table \ref{tab:modality_acc}. We can see from the table that happy faces have the highest match percentage among all affects. This means that when a happy facial animation is presented with other emotional facial animations, more than 80 percent of our raters can accurately recognize it. 
Sad and neutral faces receive a higher equal percentage because users can sometimes be uncertain when they are paired with other emotional faces. Nonetheless, all affective facial behaviors receive the average match (match + 0.5 * equal) larger than 71 percent.
On average, our affective facial parameters lead to 77\% average match, far beyond random guess (50\% average match). 
Regarding body gestures, our selection of the affective parameters leads to a slightly lower average match than the face modality, with 71.5\% average match for all affects. Boredom is the most distinguishable among all six affects, with 73\% match and 83\% average match. Sadness and surprise, however, receive 64\% and 63\% average match respectively.

\section{Main User Study} \label{sec:userstudy}
We conducted the main user study for our framework with all 4 modalities included: dialogue, voice, face, and body. We presented our stimulus with all modalities in the user study but changed the driving affect in each modality to see how the configuration of affects influences the affect perception.
We compared our affect-consistent model (AC) where all modalities share the same driving affect with the three model conditions that have an inconsistent voice (IV), inconsistent face (IF), or inconsistent body (IB) modality.
Table \ref{tab:comparison_models} lists the names of all our comparison models as well as their affect settings for all modalities. We denote the driving affect as Affect X and the inconsistent driving affect as Affect Y. 
The affect of the dialogue modality remains unchanged because the sentences are pre-scripted and all other modalities are dependent on the dialogue's content.

\begin{table}[!h]
  \caption{The affect setting for the four model conditions in our user study. AC refers to our affect consistent model. IV, IF, and IB refer to the models with inconsistent voice, face, and body modality. Affect Y is any different affect from the driving affect, Affect X.}
  \label{tab:comparison_models}
  \begin{tabular}{c|ccccl}
    \toprule
    \textbf{Model Condition} & \textbf{Dialogue} & \textbf{Voice} & \textbf{Face} & \textbf{Body} \\
    \midrule
    AC & Affect X  & Affect X  & Affect X  & Affect X \\
    IV & Affect X  & Affect Y  & Affect X  & Affect X \\
    IF & Affect X  & Affect X  & Affect Y  & Affect X \\
    IB & Affect X  & Affect X  & Affect X  & Affect Y \\
  \bottomrule
\end{tabular}
\end{table}

We rendered the multimodal responses of each model as the stimulus in our user studies.
We chose 5 dialogues from the late scenario, each containing 6 text responses associated with the 6 driving affects (Affect X in Table \ref{tab:comparison_models}). For every text response, we then generated the multimodal behaviors for the 4 comparison models. The affect-consistent model was used to generate only one sample, but the 3 affect-inconsistent models (IV, IF, and IB) were used to generate 3 samples, with different Affect Y (explained in Table \ref{tab:comparison_models}). One was neutrality and the other two were randomly selected from the remaining affects. In total, 300 video samples (5 dialogues $\times$ 6 sentences $\times$ 10 affect settings) were generated for the main user study.

We distributed our survey and collected responses from the participants  through Qualtrics \cite{qualtrics}. In the survey, each participant was first presented with a recorded video with a question displayed. We then asked the participants to answer to what extent the character's response aligned with the 6 defined affects using the 7-point Likert Scale. An answer of 1 on the scale indicates strong disagreement, 7 indicates strong agreement, and 4 is the threshold between agreement and disagreement. 
The same survey question was repeated 25 times for each participant, with videos randomly selected from the rendered responses.
In the experiment, a total of 199 participants were recruited, most of which were university undergraduates with little or no knowledge of ECAs. On average, each video was rated by 49 different participants. The survey was taken anonymously and strictly followed the university IRB rules.


\section{Results} \label{sec:results}
\newcommand\graycell{\cellcolor{gray!15}}
\begin{table*}[!t]
  \caption{The average and standard deviations of the Likert scores for the four comparison models. The columns indicate the driving affects and the rows represent the perceived affects. Bold face means the highest average score among all perceived affects. The symbol * denotes the significant decrease in the correct perception score after the removal of a consistent affect under $\alpha$ < 0.05, while $^\dagger$ means the significant decrease under $\alpha$ < 0.001. Entries with gray background mean the highest perception score does not occur when the perceived affect is the same as the driving affect. CPIN refers to the correct perception of the driving affect when the inconsistent affect is neutrality.}
  \label{tab:likert_scores}
  \begin{tabular}{c|c|c|c|c|c|c|c}
    \hline \hline
    \multirow{2}{*}{\begin{tabular}[c]{@{}c@{}}Model \\Condition\end{tabular}} & \multirow{2}{*}{\begin{tabular}[c]{@{}c@{}}Perceived \\Affect\end{tabular}} & \multicolumn{6}{c}{Driving Affect} \\
    \cline{3-8}
     &  & Neutral & Angry & Happy & Bored & Sad & Surprised \\
    \hline \hline
    \multirow{6}{*}{AC} & Neutral & \textbf{4.76 $\pm$ 1.69} & 3.12 $\pm$ 1.29 & 3.59 $\pm$ 1.53 & 4.10 $\pm$ 1.97 & 3.39 $\pm$ 1.44 & 3.78 $\pm$ 1.57 \\
                          & Angry & 3.76 $\pm$ 1.73 & \textbf{4.87 $\pm$ 1.63} & 2.97 $\pm$ 1.29 & 2.86 $\pm$ 1.13 & 2.97 $\pm$ 1.19 & 2.98 $\pm$ 1.38 \\
                          & Happy & 2.70 $\pm$ 0.79 & 3.52 $\pm$ 1.43 & \textbf{4.39 $\pm$ 1.72} & 3.38 $\pm$ 1.16 & 2.81 $\pm$ 0.95 & 3.41 $\pm$ 1.51 \\
                          & Bored & 3.55 $\pm$ 1.53 & 3.10 $\pm$ 1.13 & 3.07 $\pm$ 1.41 & \textbf{4.40 $\pm$ 1.69} & 3.05 $\pm$ 1.38 & 3.05 $\pm$ 1.28 \\
                          & Sad & 2.82 $\pm$ 1.21 & 3.03 $\pm$ 1.16 & 3.42 $\pm$ 1.41 & 3.79 $\pm$ 1.99 & \textbf{5.37 $\pm$ 1.66} & 3.63 $\pm$ 1.65 \\
                          & Surprised & 2.76 $\pm$ 0.78 & 3.27 $\pm$ 1.73 & 3.73 $\pm$ 1.63 & 3.45 $\pm$ 1.21 & 3.22 $\pm$ 1.30 & \textbf{4.11 $\pm$ 1.90} \\
    \hline \hline
    \multirow{7}{*}{IV}  & Neutral & \graycell$^{\dagger}$3.83 $\pm$ 1.76 & 3.67 $\pm$ 1.92 & \graycell\textbf{3.70 $\pm$ 1.91} & \graycell\textbf{4.30 $\pm$ 1.84} & 3.41 $\pm$ 1.55 & \graycell\textbf{3.78 $\pm$ 1.83} \\
                          & Angry & \graycell3.62 $\pm$ 1.65 & \textbf{$^{\dagger}$3.67 $\pm$ 1.67} & \graycell3.05 $\pm$ 1.44 & \graycell3.21 $\pm$ 1.57 & 2.91 $\pm$ 1.52 & \graycell3.48 $\pm$ 1.54 \\
                          & Happy & \graycell3.17 $\pm$ 1.21 & 3.16 $\pm$ 1.45 & \graycell$^{\dagger}$3.48 $\pm$ 1.54 & \graycell3.19 $\pm$ 1.29 & 2.86 $\pm$ 1.05 & \graycell3.48 $\pm$ 1.45 \\
                          & Bored & \graycell\textbf{3.95 $\pm$ 1.81} & 3.62 $\pm$ 1.71 & \graycell3.51 $\pm$ 1.67 & \graycell$^{\dagger}$3.53 $\pm$ 1.52 & 3.19 $\pm$ 1.54 & \graycell3.40 $\pm$ 1.66 \\
                          & Sad & \graycell3.54 $\pm$ 1.76 & 3.33 $\pm$ 1.60 & \graycell3.34 $\pm$ 1.59 & \graycell3.79 $\pm$ 1.91 & \textbf{*4.75 $\pm$ 1.78} & \graycell3.56 $\pm$ 1.60 \\
                          & Surprised & \graycell3.44 $\pm$ 1.37 & 3.16 $\pm$ 1.63 & \graycell3.07 $\pm$ 1.69 & \graycell3.18 $\pm$ 1.40 & 3.10 $\pm$ 1.41 & \graycell3.72 $\pm$ 1.63 \\
    \cline{2-8}
                          & CPIN & N/A & $^{\dagger}$3.72 $\pm$ 1.70  & $^{\dagger}$3.65 $\pm$ 1.63  & $^{\dagger}$3.59 $\pm$ 1.48 & *4.59 $\pm$ 1.87 & 3.70 $\pm$ 1.71 \\
    \hline \hline
    \multirow{6}{*}{IF}  & Neutral & \textbf{$^{\dagger}$3.81 $\pm$ 1.66} & 3.58 $\pm$ 1.61 & \graycell\textbf{3.66 $\pm$ 1.89} & \graycell\textbf{4.39 $\pm$ 2.07} & 3.52 $\pm$ 1.91 & \graycell\textbf{3.71 $\pm$ 1.82} \\
                          & Angry & 3.13 $\pm$ 1.58 & \textbf{$^{\dagger}$3.80 $\pm$ 1.91} & \graycell2.79 $\pm$ 1.47 & \graycell3.02 $\pm$ 1.30 & 2.93 $\pm$ 1.35 & \graycell3.36 $\pm$ 1.51 \\
                          & Happy & 3.14 $\pm$ 1.27 & 3.14 $\pm$ 1.40 & \graycell$^{\dagger}$3.49 $\pm$ 1.54 & \graycell3.29 $\pm$ 1.16 & 2.89 $\pm$ 1.01 & \graycell3.46 $\pm$ 1.37 \\
                          & Bored & 3.55 $\pm$ 1.77 & 3.33 $\pm$ 1.71 & \graycell3.33 $\pm$ 1.48 & \graycell$^{\dagger}$3.57 $\pm$ 1.69 & 3.00 $\pm$ 1.55 & \graycell3.43 $\pm$ 1.59 \\
                          & Sad & 3.47 $\pm$ 1.73 & 3.31 $\pm$ 1.44 & \graycell3.27 $\pm$ 1.45 & \graycell3.81 $\pm$ 2.06 & \textbf{$^{\dagger}$4.05 $\pm$ 1.92} & \graycell3.69 $\pm$ 2.01 \\
                          & Surprised & 3.21 $\pm$ 1.33 & 3.08 $\pm$ 1.63 & \graycell3.14 $\pm$ 1.56 & \graycell3.12 $\pm$ 1.26 & 3.01 $\pm$ 1.34 & \graycell*3.62 $\pm$ 1.69 \\
    \cline{2-8}
                          & CPIN & N/A & $^{\dagger}$3.83 $\pm$ 1.86 & $^{\dagger}$3.41 $\pm$ 1.46 & $^{\dagger}$3.63 $\pm$ 1.36 & $^{\dagger}$3.99 $\pm$ 1.89 & *3.63 $\pm$ 1.76 \\
    \hline \hline
    \multirow{6}{*}{IB}  & Neutral & \textbf{*4.18 $\pm$ 1.69} & 3.42 $\pm$ 1.57 & \graycell3.00 $\pm$ 1.60 & \graycell\textbf{3.99 $\pm$ 2.12} & 3.27 $\pm$ 1.60 & 3.58 $\pm$ 1.76 \\
                          & Angry & 3.39 $\pm$ 1.69 & \textbf{*4.39 $\pm$ 1.97} & \graycell3.16 $\pm$ 1.54 & \graycell3.04 $\pm$ 1.37 & 2.93 $\pm$ 1.27 & 3.40 $\pm$ 1.53 \\
                          & Happy & 3.09 $\pm$ 1.25 & 3.23 $\pm$ 1.37 & \graycell$^{\dagger}$3.45 $\pm$ 1.58 & \graycell3.05 $\pm$ 1.15 & 2.99 $\pm$ 1.01 & 3.22 $\pm$ 1.39 \\
                          & Bored & 3.61 $\pm$ 1.93 & 3.72 $\pm$ 1.66 & \graycell\textbf{3.49 $\pm$ 1.43} & \graycell$^{\dagger}$3.55 $\pm$ 1.78 & 3.19 $\pm$ 1.38 & 3.24 $\pm$ 1.61 \\
                          & Sad & 3.55 $\pm$ 1.74 & 3.20 $\pm$ 1.40 & \graycell2.96 $\pm$ 1.51 & \graycell3.65 $\pm$ 1.97 & \textbf{*4.69 $\pm$ 1.91} & 3.34 $\pm$ 1.87 \\
                          & Surprised & 2.99 $\pm$ 1.27 & 3.43 $\pm$ 1.58 & \graycell3.31 $\pm$ 1.87 & \graycell3.14 $\pm$ 1.19 & 3.26 $\pm$ 1.40 & \textbf{3.69 $\pm$ 1.63} \\
    \cline{2-8}
                          & CPIN & N/A & 4.46 $\pm$ 1.94 & $^{\dagger}$3.56 $\pm$ 1.44 & $^{\dagger}$3.78 $\pm$ 1.65 & 4.87 $\pm$ 1.96 & 3.86 $\pm$ 1.71 \\    
    \hline \hline
  \end{tabular}
\end{table*}

\subsection{Confusion Matrices for All Model Conditions} 

For each of the 4 model conditions, Table \ref{tab:likert_scores} reports the average and standard deviation of the Likert scores for each affect's perceived alignment with a driving affect, where the affect being compared to the driving affect is referred to as the \textit{perceived affect}. 
Each model's results are a 6$\times$6 confusion matrix, for which the diagonal entries match the perceived and driving affects. We refer to it as the perception of the driving affect or the correct perception in our results.

\subsubsection{\textbf{When all modalities are affect-consistent, participants recognize the driving affect.}}
The AC model condition receives the highest Likert scores along the diagonal entries, with all average scores greater than 4. This indicates that with consistent affects across all modalities, participants correctly recognize that there is the most alignment when the perceived and driving affects are the same
Among all the diagonal entries, participants found sadness to be the most recognizable driving affect and surprise to be the least, which indicates that the sad and surprised behaviors have the highest and lowest degree of expressiveness respectively.
Moreover, we can make observations about the non-diagonal perceived affects that may be confused with the driving affect for the AC model.
This confusion of alignment between the driving affect and perceived affect is most notable where the average Likert score of a misperceived affect is close to or greater than 4, meaning that the average participant somewhat agreed with the misalignment.
When the driving affect is happiness, the Likert score of surprise is close to 4.
When the bored behaviors are presented to the participants, the perceived neutrality and sadness are also strong.
When surprise is used as the driving affect, the perceived neutrality score is also high. Some confusions can be explained by the valence-arousal circumplex. For example, happiness and surprise are close to each other in terms of their valence arousal positions, and boredom and sadness have close proximity. However, other confusions, including the boredom-neutrality and surprise-neutrality pairs, can be the cause of low degree of expressiveness, so the participants tend to rate the emotional behaviors with higher neutrality scores.

\subsubsection{\textbf{One inconsistent modality can disrupt the recognition of almost all driving affect.}}
We can see how the removal of a consistent affect from one modality influences the perception of the driving affect, which is otherwise perceived correctly as evidenced by the AC model in Table~\ref{tab:likert_scores}. For example, the confusion matrix of the IV model shows that the inconsistent affect in the voice modality decreases the Likert scores at the diagonal entries.
Most of the Likert scores drop below 4, which suggests that the participants do not agree the perceived affects are the same as the driving affects.
We conduct the one-tailed t-test \cite{welch1947generalization} between the AC model and three other model conditions on their correct perception scores. The result is provided at the diagonal cells in Table \ref{tab:likert_scores} (denoted as * and $^\dagger$). There are significant decreases in Likert scores after the removal of consistent affect in voice and face modality for the correct perception of all affects. However, there is no significant decrease after the removal of a consistent affect in body modality for surprise perception. 
The result also implies that an inconsistent affect in the voice and face modalities leads to more statistically significant decreases than the body modality in the correct perception scores.

\subsubsection{\textbf{Some affects are more resilient than others to the inconsistency of one modality.}} When a consistent affect is removed from any of the three modalities, we see that the two driving affects, anger and sadness, can still be recognized by participants. Neutrality and surprise are less resilient because the perception of the same driving affect is the highest after the removal of each of the two affects in the body modality. When a happy or bored consistent affect is removed from any of the three modalities, the correct perception of the two affects is largely influenced. In fact, we can see a link between the decrease in the correct perception of driving affect and the increase in the perception of neutrality. The entries highlighted in gray at Table \ref{tab:likert_scores} indicate that the perceived neutrality score also increases for those irresilient driving affects.   
The three driving affects, happiness, boredom, and surprise are even perceived as more neutral for IV and IF models. 
We attribute the decrease in the correct perception of driving affect as well as the increase in perceived neutrality after the removal of a consistent affect to the cause of emotion dilution, when the expressiveness of the multimodal behavior is discounted by affect inconsistency.
Overall, we observe that the removal of a consistent affect in voice (IV) and face (IF) modalities increases the perception of neutrality more than the removal in the body (IB) modality.
This also implies that the emotion dilution issue is more obvious when either the voice or the face modality has an inconsistent affect. 

\subsubsection{\textbf{A modality without emotion conditioning can decrease the perception of the driving affect.}}
We have mentioned in Section \ref{sec:conceptual framework} that emotion conditioning can be helpful as it mitigates the emotion dilution issue during the sequential executions for multimodal behaviors. We regard the stimulus with the inconsistent neutrality affect in the IV, IF, and IB model conditions as the behaviors generated without emotion conditioning in the voice, face, and body modalities respectively. Each 'CPIN' row in Table \ref{tab:likert_scores} reports the average and standard deviation of the Likert scores for the correct perception of all driving affects. Without emotion conditioning in one modality, the perception of the driving affect decreases, when compared with the AC model with all emotion-conditioned modalities. We also notice that emotion dilution is more obvious when the correct perception score is much more influenced by the unconditioned modality. This suggests that when the emotion conditioning is removed from one modality, the generated behaviors in that modality are perceived as less expressive and more neutral, which then decreases the correct perception of affects when other affective modalities are combined. 

\subsection{Interaction and Main Effects on Perception}
We conduct the 3-way ANOVA test \cite{judd2017data} to analyze the effect of the three independent variables, \textit{model}, \textit{driving affect}, and \textit{perceived affect}, on the perception Likert scores and we apply Tukey HSD \cite{abdi2010tukey} for the post hoc tests. There are 4 conditions in the model variable and 6 conditions each for the two remaining variables. 
The result shows that there is a statistically significant interaction between the effects of \textit{model}, \textit{driving affect}, and \textit{perceived affect} on the perception Likert score, with F(75, 12444) = 2.105 and \textit{p} < 0.001.

Specifically, we observe a main effect on \textit{perceived affect}, with F(5, 12444) = 28.040 and \textit{p} < 0.001. The post hoc test indicates that across all conditions in the \textit{model} and \textit{driving affect} variables, the perception scores of neutrality and sadness are significantly higher than the scores of anger, happiness, boredom, and surprise, and the boredom perception score is significantly higher than happiness.
This implies that the participants tend to rate the stimulus with higher Likert scores in neutrality and sadness. The reason behind the higher sadness rating is that it receives the highest when the driving affect is sadness. For neutrality, we can observe from Table \ref{tab:likert_scores} that neutrality is often misperceived as the highest for all affect inconsistent models.
Across all driving affects, there is a simple main effect of the \textit{perceived affect} variable on the Likert score for every model condition. The perception of the neutrality affect under AC model is statistically more significant than the perception of boredom, surprise, anger, and happiness, while the neutrality perception is much more significant than the perception of all other affects for all three affect inconsistent models (IV, IF, and IB). This shows that the emotion dilution is more obvious when the affects are inconsistent. 

We also observe a main interaction effect between the \textit{driving affect} and \textit{perceived affect}, with F(25, 12444) = 24.328 and p < 0.001. Ideally, when the driving affect matches the perceived affect, the Likert score should be significantly higher than other driving affects.
For all 6 conditions in the \textit{perceived affect} variable, all the conditions in the \textit{driving affect} variable with the same affect condition show statistically significant differences from other driving affects with dissimilar conditions, except one pair, neutrality, and boredom. 
The reason can be seen from Table \ref{tab:likert_scores} that when the perceived affect is neutrality, the boredom driving affect is rated relatively high for the three affect inconsistent models. 
We observe a similar simple effect of the \textit{driving affect} variable on the Likert score when the AC model is considered. All driving affects that are the same as the perceived affect are rated significantly higher than the driving affects that are different from the perceived affects. However, when the perceived affect is surprise, the Likert score for the surprise driving affect is not rated significantly different from happiness. According to Table \ref{tab:likert_scores}, the average Likert score for happiness driving affect and perceived surprise is 3.73 and the perceived surprise score for surprise driving affect is 4.11. This suggests that the surprised multimodal behavior is not significantly different from the happy behavior in terms of the perception of surprise.

As we observe the interaction effect of \textit{driving affect * perceived affect}, we are also interested in the comparisons of the \textit{perceived affects} on the Likert score under each \textit{model} condition. Ideally when the perceived affect is the same as the driving affect, the Likert score should be significantly higher than other perceived affects. 
We find most comparison pairs follow the above expectation for AC model. However, when the driving affect is boredom, the perception of boredom is not significantly different from sadness. The result suggests that there is confusion between the perception of boredom and sadness when the bored behavior is provided. We believe this is due to the proximity of boredom and sadness on their valence-arousal values \cite{russell1980circumplex}. When the driving affect is surprise or boredom, we find no significant differences between the perception of the same affect and neutrality, which indicates that these generated emotional behaviors are not as  expressive as others.



\subsection{Effects on Correct Perception of Affects} 

We are also interested in the interaction of the model and the driving affect on the correct perception of Likert score. The correct perception of an expressive behavior means the perception of the same affect as the driving affect. In other words, all the diagonal entries in Table \ref{tab:likert_scores} are considered. We further conduct the 2-way ANOVA test to report the effect of the two independent variables, \textit{model} and \textit{driving affect}, on the correct affect perception in Likert scores. 
The result shows that there is no significant interaction of the two independent variables on the correct perception, with F(15, 2815) = 1.640 and \textit{p} < 0.056. However, we observe the main effects of \textit{model} and \textit{driving affect}, with F(3, 2815) = 27.773, \textit{p} < 0.001 and F(5, 2815) = 21.578, \textit{p} < 0.001 respectively. We apply Tukey HSD for the post hoc test for the comparisons of the conditions in the two variables.

\begin{figure}[!t] 
  \centering
  \includegraphics[trim={1.0cm 0.1cm 1.2cm 0.5cm},clip, width=\linewidth]{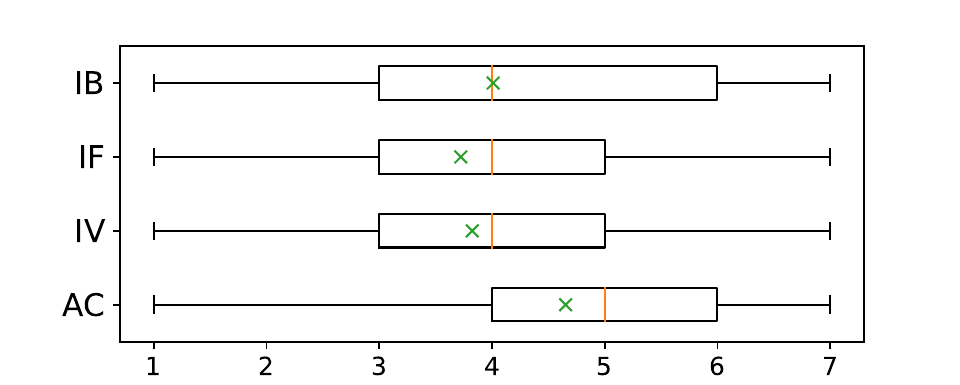}
  \caption{The box plot of the correct perception score for the four model conditions. 'X' denotes the average score.}
  \label{fig:boxplot}
\end{figure}

\subsubsection{Difference in Models}
The differences between the AC model and the IV, IF, and IB models across all driving affects represents the discrepancy in the correct affect perception between an affect-consistent behavior and an inconsistent behavior. The result indicates that across all driving affects, the AC model receives a significantly higher Likert score than the IV, IF, and IB  model conditions, with average differences = 0.822, 0.928, 0.660 and \textit{p} < 0.001. 
Adding a consistent affect to one of the voice, face, and body modality can significantly improve the perception of the driving affect. Specifically, the average score differences in voice and face modality are higher than the body modality, which suggests that the voice and face modality are more important than body when expressing affects. We can observe the same finding from Figure~\ref{fig:boxplot} that IB model condition has higher average score than IV and IF, and still receives a similar 75th percentile as AC model condition. Our findings are aligned with the previous studies \cite{sonlu2021conversational, ennis2013emotion, clavel2009combining} that all modalities do contribute to affect perception and combining the expressive modalities is more helpful. The perception, however, is not mainly judged by body expressions.

\subsubsection{Difference in Driving Affects}
The comparison of the conditions in \textit{driving affect} on the correct perception score reflects how easily the affects can be recognized by the participants. Our result indicates that when the driving affect is sadness, the correct perception score is significantly higher than all other affects, across all \textit{model} conditions. The three affects, happiness, boredom, and surprise, however, are hard to recognize as they receive relatively lower scores than the other three affects. We observe that the removal of the three consistent affects in a modality would largely decrease their correct perception scores, while the other three are somewhat more resilient. The low expressiveness of the three emotional behaviors could be the limitation of our collected methods.
Furthermore, when looking specifically at every model condition, we can still tell that sadness is the most recognizable among all the affects, as it receives the highest score for correct perception. However, we can only see significant differences in the correct perception between the two pairs, sadness-happiness, and sadness-surprise, where the affects have opposite valence and arousal values. This shows that the sad face is a major indicator of the recognition of sadness.

\section{Discussion} \label{sec:discussion}

The results of our user study indicate that affect consistency maximizes the perception of the driving affect and making a single modality's affect inconsistent decreases the perception of the correct affect.
The conclusion is supported by previous studies \cite{clavel2009combining, ennis2013emotion}, which investigated the effect that the presence of face and body modalities had on perception. They found that both modalities individually contribute to affect perception and that the perception is maximized when both modalities are present. Furthermore, our experimental analysis suggests that the voice and face modalities contribute more than the body modality to the perception of affects. This finding accords with prior works \citep{sonlu2021conversational, ennis2013emotion} regarding the perception of emotions for virtual characters, although the experimental designs are different.
While their stimuli contain all combinations of the presence of each modality with consistent affect, ours include all modalities with both consistent and inconsistent affects.
The differences in the modalities in our study need to be analyzed by comparing the differences between the affect consistent model and inconsistent models. 
Nonetheless, our comparisons of modalities can be done in a more natural conversation scenario where all modalities are present.

From the statistical analysis of the main and simple effects of \textit{perceived affect} variable, we can observe that overall, neutrality is strongly perceived, despite it not being the driving affect. 
The observation still holds true regardless of the \textit{model} condition, but the perception of neutrality is stronger when a modality has an inconsistent affect. Our study shows that the emotion dilution occurs when the one modality has an inconsistent affect and the participants tend to rate a higher neutrality perception score. The issue, however, is mitigated under the AC model condition where all the modalities are generated with consistent emotion conditioning. Evidenced by the experimental results and analyses, we can conclude that our conceptual framework for ECAs with multimodal emotion conditioning and affect consistency successfully addresses emotion dilution and enhances the correct perception of affect. 

The correlations of the affects are investigated in our study through the analyses of the interaction effect between the \textit{driving affect} and \textit{perceived affect} as well as the difference in the effect of the \textit{driving affects} on the correct perception. We do observe certain links that explain the correlations of affects in the confusion matrix with the valence-arousal circumplex. For example, boredom can be confused with sadness, and happiness is correlated with surprise. 
However, we do not observe that the body modality helps the participants discriminate emotions based on their valence values, as suggested by the previous study \cite{aviezer2012body}. 
There are some confusions that cannot be explained by their valence-arousal values. For instance, the neutrality perception scores are also high when the driving affects are boredom and surprise, which indicates that the two emotional behaviors are not as expressive as other affective behaviors. The reason could result from the limitation of the collected data-driven models. Currently, the expressiveness of the framework is dependent on the methods we use to generate the modalities. We can see the dependence from our results that sadness is the most recognizable affect among all, but the previous works \cite{clavel2009combining, ennis2013emotion} on the perception of virtual characters revealed that anger and happiness are more easily recognized, and sadness is the most difficult to tell. We acknowledge a gap in expressiveness between the real human behaviors that were used as stimulus by previous studies and the generated human behaviors in our experiments. Nonetheless, the affect-consistent framework is able to generate emotional behaviors with increased correct perception.

Overall, our framework combines the existing methods for the voice, face, and body modalities and creates a shared affect space for generating affect consistent behaviors. 
We could see several limitations. 
First, the quality of the affect consistent framework is dependent on those methods. For instance, the use of gesture generation from audio and text by the framework could potentially lead to a loss of communication efficacy when compared with rule-based methods \cite{ravenet2018automating, bergmann2009increasing, wagner2014gesture}. According to the classification of gesture types proposed by \cite{mcneill1992hand}, our generative approach could only generate beat and metaphoric gestures from the input audio and text.   
Second, the mappings of the 6 affects and the latent parameters are only specific to our implementation, which decreases the generalizability of the proposed framework.
However, the methods can be easily substituted and the mappings can be obtained in different ways accordingly. For example, the affect mappings can be heuristics-based, manually selected and verified through user study, or achieved via supervised training. 
The framework itself can be improved as a better method for modality generation is available. 
Still, the importance of the two properties, emotion conditioning and affect consistency, holds true. 

On the other hand, when it comes to real-time interaction, our framework does require longer process time compared with pre-authored behaviors. The latency comes from the interdependence of the modalities, so all the modalities cannot be generated in parallel. The incorporation of separate generative methods into the framework also introduces additional latency as the generated facial and body animations have to be retargeted to the same character at runtime.


\section{Conclusion}
In this work, we propose a conceptual framework, \framework, with affect-consistent multimodal behaviors for ECAs that aims to enhance the user's perception of affects. 
We conduct the main user study for the evaluation of our framework. The result indicates that the multimodal behavior with consistent affect receives the highest correct perception score and removing a consistent affect from the voice, face, and body modalities can significantly decrease the perception of the driving affect. Our statistical analysis also suggests that emotion conditioning and affect consistency are helpful for mitigating the emotion dilution issue.


\begin{acks}
The research was supported in part by NSF awards: IIS-1703883, IIS-1955404, IIS-1955365, RETTL-2119265, and EAGER-2122119.
This material is based upon work supported by the U.S. Department of Homeland Security\footnote{Disclaimer. The views and conclusions contained in this document are those of the authors and should not be interpreted as necessarily representing the official policies, either expressed or implied, of the U.S. Department of Homeland Security.} under Grant Award Number 22STESE00001 01 01. This publication is based upon work supported by King Fahd University of Petroleum \& Minerals. Author(s) at KFUPM acknowledge the Interdisciplinary Research Center for Intelligent Secure Systems for the support received under Grant Number INSS2305.
\end{acks}

\bibliographystyle{ACM-Reference-Format}
\bibliography{main}


\begin{thebibliography}{64}


\ifx \showCODEN    \undefined \def \showCODEN     #1{\unskip}     \fi
\ifx \showDOI      \undefined \def \showDOI       #1{#1}\fi
\ifx \showISBNx    \undefined \def \showISBNx     #1{\unskip}     \fi
\ifx \showISBNxiii \undefined \def \showISBNxiii  #1{\unskip}     \fi
\ifx \showISSN     \undefined \def \showISSN      #1{\unskip}     \fi
\ifx \showLCCN     \undefined \def \showLCCN      #1{\unskip}     \fi
\ifx \shownote     \undefined \def \shownote      #1{#1}          \fi
\ifx \showarticletitle \undefined \def \showarticletitle #1{#1}   \fi
\ifx \showURL      \undefined \def \showURL       {\relax}        \fi
\providecommand\bibfield[2]{#2}
\providecommand\bibinfo[2]{#2}
\providecommand\natexlab[1]{#1}
\providecommand\showeprint[2][]{arXiv:#2}

\bibitem[Abdi and Williams(2010)]%
        {abdi2010tukey}
\bibfield{author}{\bibinfo{person}{Herv{\'e} Abdi} {and}
  \bibinfo{person}{Lynne~J Williams}.} \bibinfo{year}{2010}\natexlab{}.
\newblock \showarticletitle{Tukey’s honestly significant difference (HSD)
  test}.
\newblock \bibinfo{journal}{\emph{Encyclopedia of research design}}
  \bibinfo{volume}{3}, \bibinfo{number}{1} (\bibinfo{year}{2010}),
  \bibinfo{pages}{1--5}.
\newblock


\bibitem[Alexanderson et~al\mbox{.}(2020)]%
        {alexanderson2020style}
\bibfield{author}{\bibinfo{person}{Simon Alexanderson},
  \bibinfo{person}{Gustav~Eje Henter}, \bibinfo{person}{Taras Kucherenko},
  {and} \bibinfo{person}{Jonas Beskow}.} \bibinfo{year}{2020}\natexlab{}.
\newblock \showarticletitle{Style-Controllable Speech-Driven Gesture Synthesis
  Using Normalising Flows}. In \bibinfo{booktitle}{\emph{Computer Graphics
  Forum}}, Vol.~\bibinfo{volume}{39}. Wiley Online Library,
  \bibinfo{pages}{487--496}.
\newblock


\bibitem[Aviezer et~al\mbox{.}(2012)]%
        {aviezer2012body}
\bibfield{author}{\bibinfo{person}{Hillel Aviezer}, \bibinfo{person}{Yaacov
  Trope}, {and} \bibinfo{person}{Alexander Todorov}.}
  \bibinfo{year}{2012}\natexlab{}.
\newblock \showarticletitle{Body cues, not facial expressions, discriminate
  between intense positive and negative emotions}.
\newblock \bibinfo{journal}{\emph{Science}} \bibinfo{volume}{338},
  \bibinfo{number}{6111} (\bibinfo{year}{2012}), \bibinfo{pages}{1225--1229}.
\newblock


\bibitem[Badiee and Kaufman(2015)]%
        {badiee2015design}
\bibfield{author}{\bibinfo{person}{Farnaz Badiee} {and} \bibinfo{person}{David
  Kaufman}.} \bibinfo{year}{2015}\natexlab{}.
\newblock \showarticletitle{Design evaluation of a simulation for teacher
  education}.
\newblock \bibinfo{journal}{\emph{Sage Open}} \bibinfo{volume}{5},
  \bibinfo{number}{2} (\bibinfo{year}{2015}),
  \bibinfo{pages}{2158244015592454}.
\newblock


\bibitem[Bergmann and Kopp(2009)]%
        {bergmann2009increasing}
\bibfield{author}{\bibinfo{person}{Kirsten Bergmann} {and}
  \bibinfo{person}{Stefan Kopp}.} \bibinfo{year}{2009}\natexlab{}.
\newblock \showarticletitle{Increasing the expressiveness of virtual agents:
  autonomous generation of speech and gesture for spatial description tasks}.
  In \bibinfo{booktitle}{\emph{Proceedings of The 8th International Conference
  on Autonomous Agents and Multiagent Systems-Volume 1}}.
  \bibinfo{pages}{361--368}.
\newblock


\bibitem[Bhattacharya et~al\mbox{.}(2021a)]%
        {bhattacharya2021speech2affectivegestures}
\bibfield{author}{\bibinfo{person}{Uttaran Bhattacharya},
  \bibinfo{person}{Elizabeth Childs}, \bibinfo{person}{Nicholas Rewkowski},
  {and} \bibinfo{person}{Dinesh Manocha}.} \bibinfo{year}{2021}\natexlab{a}.
\newblock \showarticletitle{Speech2AffectiveGestures: Synthesizing Co-Speech
  Gestures with Generative Adversarial Affective Expression Learning}. In
  \bibinfo{booktitle}{\emph{Proceedings of the 29th ACM International
  Conference on Multimedia}} \emph{(\bibinfo{series}{MM '21})}.
  \bibinfo{publisher}{Association for Computing Machinery},
  \bibinfo{address}{New York, NY, USA}.
\newblock


\bibitem[Bhattacharya et~al\mbox{.}(2021b)]%
        {bhattacharya2021text2gestures}
\bibfield{author}{\bibinfo{person}{Uttaran Bhattacharya},
  \bibinfo{person}{Nicholas Rewkowski}, \bibinfo{person}{Abhishek Banerjee},
  \bibinfo{person}{Pooja Guhan}, \bibinfo{person}{Aniket Bera}, {and}
  \bibinfo{person}{Dinesh Manocha}.} \bibinfo{year}{2021}\natexlab{b}.
\newblock \showarticletitle{Text2Gestures: A Transformer-Based Network for
  Generating Emotive Body Gestures for Virtual Agents** This work has been
  supported in part by ARO Grants W911NF1910069 and W911NF1910315, and Intel.
  Code and additional materials available at: https://gamma. umd. edu/t2g}. In
  \bibinfo{booktitle}{\emph{2021 IEEE Virtual Reality and 3D User Interfaces
  (VR)}}. IEEE, \bibinfo{pages}{1--10}.
\newblock


\bibitem[Blomsma et~al\mbox{.}(2020)]%
        {blomsma2020intrapersonal}
\bibfield{author}{\bibinfo{person}{Pieter~A Blomsma}, \bibinfo{person}{Guido~M
  Linders}, \bibinfo{person}{Julija Vaitonyte}, {and} \bibinfo{person}{Max~M
  Louwerse}.} \bibinfo{year}{2020}\natexlab{}.
\newblock \showarticletitle{Intrapersonal dependencies in multimodal behavior}.
  In \bibinfo{booktitle}{\emph{Proceedings of the 20th ACM International
  Conference on Intelligent Virtual Agents}}. \bibinfo{pages}{1--8}.
\newblock


\bibitem[Casas et~al\mbox{.}(2021)]%
        {casas2021enhancing}
\bibfield{author}{\bibinfo{person}{Jacky Casas}, \bibinfo{person}{Timo Spring},
  \bibinfo{person}{Karl Daher}, \bibinfo{person}{Elena Mugellini},
  \bibinfo{person}{Omar~Abou Khaled}, {and} \bibinfo{person}{Philippe
  Cudr{\'e}-Mauroux}.} \bibinfo{year}{2021}\natexlab{}.
\newblock \showarticletitle{Enhancing conversational agents with empathic
  abilities}. In \bibinfo{booktitle}{\emph{Proceedings of the 21st ACM
  International Conference on Intelligent Virtual Agents}}.
  \bibinfo{pages}{41--47}.
\newblock


\bibitem[Castillo et~al\mbox{.}(2018)]%
        {castillo2018personality}
\bibfield{author}{\bibinfo{person}{Susana Castillo}, \bibinfo{person}{Philipp
  Hahn}, \bibinfo{person}{Katharina Legde}, {and} \bibinfo{person}{Douglas~W
  Cunningham}.} \bibinfo{year}{2018}\natexlab{}.
\newblock \showarticletitle{Personality analysis of embodied conversational
  agents}. In \bibinfo{booktitle}{\emph{Proceedings of the 18th International
  Conference on Intelligent Virtual Agents}}. \bibinfo{pages}{227--232}.
\newblock


\bibitem[Chang(2020)]%
        {chang2020transfer}
\bibfield{author}{\bibinfo{person}{Che-Jui Chang}.}
  \bibinfo{year}{2020}\natexlab{}.
\newblock \bibinfo{title}{Transfer Learning from Monolingual ASR to
  Transcription-free Cross-lingual Voice Conversion}.
\newblock
\newblock
\urldef\tempurl%
\url{https://doi.org/10.48550/ARXIV.2009.14668}
\showDOI{\tempurl}


\bibitem[Chang et~al\mbox{.}(2022a)]%
        {chang2022ivi}
\bibfield{author}{\bibinfo{person}{Che-Jui Chang}, \bibinfo{person}{Sen Zhang},
  {and} \bibinfo{person}{Mubbasir Kapadia}.} \bibinfo{year}{2022}\natexlab{a}.
\newblock \showarticletitle{The IVI Lab entry to the GENEA Challenge 2022--A
  Tacotron2 based method for co-speech gesture generation with
  locality-constraint attention mechanism}. In
  \bibinfo{booktitle}{\emph{Proceedings of the 2022 International Conference on
  Multimodal Interaction}}. \bibinfo{pages}{784--789}.
\newblock


\bibitem[Chang et~al\mbox{.}(2022b)]%
        {chang2022disentangle}
\bibfield{author}{\bibinfo{person}{Che-Jui Chang}, \bibinfo{person}{Long Zhao},
  \bibinfo{person}{Sen Zhang}, {and} \bibinfo{person}{Mubbasir Kapadia}.}
  \bibinfo{year}{2022}\natexlab{b}.
\newblock \showarticletitle{Disentangling audio content and emotion with
  adaptive instance normalization for expressive facial animation synthesis}.
\newblock \bibinfo{journal}{\emph{Computer Animation and Virtual Worlds}}
  \bibinfo{volume}{33}, \bibinfo{number}{3-4} (\bibinfo{year}{2022}),
  \bibinfo{pages}{e2076}.
\newblock
\urldef\tempurl%
\url{https://doi.org/10.1002/cav.2076}
\showDOI{\tempurl}
\showeprint{https://onlinelibrary.wiley.com/doi/pdf/10.1002/cav.2076}


\bibitem[Chen et~al\mbox{.}(2020)]%
        {chen2020dynamic}
\bibfield{author}{\bibinfo{person}{Chaona Chen}, \bibinfo{person}{Oliver~GB
  Garrod}, \bibinfo{person}{Philippe~G Schyns}, {and}
  \bibinfo{person}{Rachael~E Jack}.} \bibinfo{year}{2020}\natexlab{}.
\newblock \showarticletitle{Dynamic Face Movement Texture Enhances the
  Perceived Realism of Facial Expressions of Emotion}. In
  \bibinfo{booktitle}{\emph{Proceedings of the 20th ACM International
  Conference on Intelligent Virtual Agents}}. \bibinfo{pages}{1--3}.
\newblock


\bibitem[Clavel et~al\mbox{.}(2009)]%
        {clavel2009combining}
\bibfield{author}{\bibinfo{person}{C{\'e}line Clavel}, \bibinfo{person}{Justine
  Plessier}, \bibinfo{person}{Jean-Claude Martin}, \bibinfo{person}{Laurent
  Ach}, {and} \bibinfo{person}{Benoit Morel}.} \bibinfo{year}{2009}\natexlab{}.
\newblock \showarticletitle{Combining facial and postural expressions of
  emotions in a virtual character}. In \bibinfo{booktitle}{\emph{International
  Workshop on Intelligent Virtual Agents}}. Springer,
  \bibinfo{pages}{287--300}.
\newblock


\bibitem[Colombo et~al\mbox{.}(2019)]%
        {colombo2019affect-driven}
\bibfield{author}{\bibinfo{person}{Pierre Colombo}, \bibinfo{person}{Wojciech
  Witon}, \bibinfo{person}{Ashutosh Modi}, \bibinfo{person}{James Kennedy},
  {and} \bibinfo{person}{Mubbasir Kapadia}.} \bibinfo{year}{2019}\natexlab{}.
\newblock \showarticletitle{Affect-Driven Dialog Generation}. In
  \bibinfo{booktitle}{\emph{North American Chapter of the Association for
  Computational Linguistics: Human Language Technologies, NAACL-HLT}}
  (2019-01-01). \bibinfo{pages}{3734--3743}.
\newblock
\urldef\tempurl%
\url{https://aclweb.org/anthology/papers/N/N19/N19-1374/}
\showURL{%
\tempurl}


\bibitem[Delamarre et~al\mbox{.}(2019)]%
        {delamarre2019aimer}
\bibfield{author}{\bibinfo{person}{Alban Delamarre},
  \bibinfo{person}{C{\'e}dric Buche}, {and} \bibinfo{person}{Christine
  Lisetti}.} \bibinfo{year}{2019}\natexlab{}.
\newblock \showarticletitle{Aimer: Appraisal interpersonal model of emotion
  regulation, affective virtual students to support teachers training}. In
  \bibinfo{booktitle}{\emph{Proceedings of the 19th ACM International
  Conference on Intelligent Virtual Agents}}. \bibinfo{pages}{182--184}.
\newblock


\bibitem[DiPaola and Yal{\c{c}}in(2019)]%
        {dipaola2019multi}
\bibfield{author}{\bibinfo{person}{Steve DiPaola} {and}
  \bibinfo{person}{{\"O}zge~Nilay Yal{\c{c}}in}.}
  \bibinfo{year}{2019}\natexlab{}.
\newblock \showarticletitle{A multi-layer artificial intelligence and sensing
  based affective conversational embodied agent}. In
  \bibinfo{booktitle}{\emph{2019 8th International Conference on Affective
  Computing and Intelligent Interaction Workshops and Demos (ACIIW)}}. IEEE,
  \bibinfo{pages}{91--92}.
\newblock


\bibitem[Durupinar et~al\mbox{.}(2016)]%
        {durupinar2016perform}
\bibfield{author}{\bibinfo{person}{Funda Durupinar}, \bibinfo{person}{Mubbasir
  Kapadia}, \bibinfo{person}{Susan Deutsch}, \bibinfo{person}{Michael Neff},
  {and} \bibinfo{person}{Norman~I Badler}.} \bibinfo{year}{2016}\natexlab{}.
\newblock \showarticletitle{Perform: Perceptual approach for adding ocean
  personality to human motion using laban movement analysis}.
\newblock \bibinfo{journal}{\emph{ACM Transactions on Graphics (TOG)}}
  \bibinfo{volume}{36}, \bibinfo{number}{1} (\bibinfo{year}{2016}),
  \bibinfo{pages}{1--16}.
\newblock


\bibitem[Ekman(1999)]%
        {ekman1999basic}
\bibfield{author}{\bibinfo{person}{Paul Ekman}.}
  \bibinfo{year}{1999}\natexlab{}.
\newblock \showarticletitle{Basic emotions}.
\newblock \bibinfo{journal}{\emph{Handbook of cognition and emotion}}
  \bibinfo{volume}{98}, \bibinfo{number}{45-60} (\bibinfo{year}{1999}),
  \bibinfo{pages}{16}.
\newblock


\bibitem[Engine(2021)]%
        {metahuman2021}
\bibfield{author}{\bibinfo{person}{Unreal Engine}.}
  \bibinfo{year}{2021}\natexlab{}.
\newblock \bibinfo{title}{MetaHuman Creator}.
\newblock
\newblock


\bibitem[Ennis et~al\mbox{.}(2013)]%
        {ennis2013emotion}
\bibfield{author}{\bibinfo{person}{Cathy Ennis}, \bibinfo{person}{Ludovic
  Hoyet}, \bibinfo{person}{Arjan Egges}, {and} \bibinfo{person}{Rachel
  McDonnell}.} \bibinfo{year}{2013}\natexlab{}.
\newblock \showarticletitle{Emotion capture: Emotionally expressive characters
  for games}.
\newblock In \bibinfo{booktitle}{\emph{Proceedings of motion on games}}.
  \bibinfo{pages}{53--60}.
\newblock


\bibitem[Falk et~al\mbox{.}(2018)]%
        {falk2018pica}
\bibfield{author}{\bibinfo{person}{Jessica Falk}, \bibinfo{person}{Steven
  Poulakos}, \bibinfo{person}{Mubbasir Kapadia}, {and}
  \bibinfo{person}{Robert~W Sumner}.} \bibinfo{year}{2018}\natexlab{}.
\newblock \showarticletitle{Pica: Proactive intelligent conversational agent
  for interactive narratives}. In \bibinfo{booktitle}{\emph{Proceedings of the
  18th International Conference on Intelligent Virtual Agents}}.
  \bibinfo{pages}{141--146}.
\newblock


\bibitem[Ferstl et~al\mbox{.}(2020)]%
        {ferstl2020understanding}
\bibfield{author}{\bibinfo{person}{Ylva Ferstl}, \bibinfo{person}{Michael
  Neff}, {and} \bibinfo{person}{Rachel McDonnell}.}
  \bibinfo{year}{2020}\natexlab{}.
\newblock \showarticletitle{Understanding the predictability of gesture
  parameters from speech and their perceptual importance}. In
  \bibinfo{booktitle}{\emph{Proceedings of the 20th ACM International
  Conference on Intelligent Virtual Agents}}. \bibinfo{pages}{1--8}.
\newblock


\bibitem[Ferstl et~al\mbox{.}(2021a)]%
        {ferstl2021expressgesture}
\bibfield{author}{\bibinfo{person}{Ylva Ferstl}, \bibinfo{person}{Michael
  Neff}, {and} \bibinfo{person}{Rachel McDonnell}.}
  \bibinfo{year}{2021}\natexlab{a}.
\newblock \showarticletitle{ExpressGesture: Expressive gesture generation from
  speech through database matching}.
\newblock \bibinfo{journal}{\emph{Computer Animation and Virtual Worlds}}
  \bibinfo{volume}{32}, \bibinfo{number}{3-4} (\bibinfo{year}{2021}),
  \bibinfo{pages}{e2016}.
\newblock


\bibitem[Ferstl et~al\mbox{.}(2021b)]%
        {ferstl2021human}
\bibfield{author}{\bibinfo{person}{Ylva Ferstl}, \bibinfo{person}{Sean Thomas},
  \bibinfo{person}{C{\'e}dric Guiard}, \bibinfo{person}{Cathy Ennis}, {and}
  \bibinfo{person}{Rachel McDonnell}.} \bibinfo{year}{2021}\natexlab{b}.
\newblock \showarticletitle{Human or Robot? Investigating voice, appearance and
  gesture motion realism of conversational social agents}. In
  \bibinfo{booktitle}{\emph{Proceedings of the 21st ACM International
  Conference on Intelligent Virtual Agents}}. \bibinfo{pages}{76--83}.
\newblock


\bibitem[Ginosar et~al\mbox{.}(2019)]%
        {ginosar2019learning}
\bibfield{author}{\bibinfo{person}{Shiry Ginosar}, \bibinfo{person}{Amir Bar},
  \bibinfo{person}{Gefen Kohavi}, \bibinfo{person}{Caroline Chan},
  \bibinfo{person}{Andrew Owens}, {and} \bibinfo{person}{Jitendra Malik}.}
  \bibinfo{year}{2019}\natexlab{}.
\newblock \showarticletitle{Learning individual styles of conversational
  gesture}. In \bibinfo{booktitle}{\emph{Proceedings of the IEEE/CVF Conference
  on Computer Vision and Pattern Recognition}}. \bibinfo{pages}{3497--3506}.
\newblock


\bibitem[Habibie et~al\mbox{.}(2021)]%
        {habibie2021learning}
\bibfield{author}{\bibinfo{person}{Ikhsanul Habibie}, \bibinfo{person}{Weipeng
  Xu}, \bibinfo{person}{Dushyant Mehta}, \bibinfo{person}{Lingjie Liu},
  \bibinfo{person}{Hans-Peter Seidel}, \bibinfo{person}{Gerard Pons-Moll},
  \bibinfo{person}{Mohamed Elgharib}, {and} \bibinfo{person}{Christian
  Theobalt}.} \bibinfo{year}{2021}\natexlab{}.
\newblock \showarticletitle{Learning speech-driven 3d conversational gestures
  from video}. In \bibinfo{booktitle}{\emph{Proceedings of the 21st ACM
  International Conference on Intelligent Virtual Agents}}.
  \bibinfo{pages}{101--108}.
\newblock


\bibitem[Hoegen et~al\mbox{.}(2019)]%
        {hoegen2019end}
\bibfield{author}{\bibinfo{person}{Rens Hoegen}, \bibinfo{person}{Deepali
  Aneja}, \bibinfo{person}{Daniel McDuff}, {and} \bibinfo{person}{Mary
  Czerwinski}.} \bibinfo{year}{2019}\natexlab{}.
\newblock \showarticletitle{An end-to-end conversational style matching agent}.
  In \bibinfo{booktitle}{\emph{Proceedings of the 19th ACM International
  Conference on Intelligent Virtual Agents}}. \bibinfo{pages}{111--118}.
\newblock


\bibitem[IBM(2015)]%
        {ibm2015}
\bibfield{author}{\bibinfo{person}{IBM}.} \bibinfo{year}{2015}\natexlab{}.
\newblock \bibinfo{title}{IBM Text to Speech}.
\newblock \bibinfo{howpublished}{\url{https://www.ibm.com/watson}}.
\newblock
\newblock
\shownote{Accessed: 2022-03-05}.


\bibitem[Ishii et~al\mbox{.}(2020)]%
        {ishii2020impact}
\bibfield{author}{\bibinfo{person}{Ryo Ishii}, \bibinfo{person}{Chaitanya
  Ahuja}, \bibinfo{person}{Yukiko~I Nakano}, {and}
  \bibinfo{person}{Louis-Philippe Morency}.} \bibinfo{year}{2020}\natexlab{}.
\newblock \showarticletitle{Impact of personality on nonverbal behavior
  generation}. In \bibinfo{booktitle}{\emph{Proceedings of the 20th ACM
  International Conference on Intelligent Virtual Agents}}.
  \bibinfo{pages}{1--8}.
\newblock


\bibitem[Janghorbani et~al\mbox{.}(2019)]%
        {janghorbani2019domain}
\bibfield{author}{\bibinfo{person}{Sepehr Janghorbani},
  \bibinfo{person}{Ashutosh Modi}, \bibinfo{person}{Jakob Buhmann}, {and}
  \bibinfo{person}{Mubbasir Kapadia}.} \bibinfo{year}{2019}\natexlab{}.
\newblock \showarticletitle{Domain authoring assistant for intelligent virtual
  agents}.
\newblock \bibinfo{journal}{\emph{arXiv preprint arXiv:1904.03266}}
  (\bibinfo{year}{2019}).
\newblock


\bibitem[Judd et~al\mbox{.}(2017)]%
        {judd2017data}
\bibfield{author}{\bibinfo{person}{Charles~M Judd}, \bibinfo{person}{Gary~H
  McClelland}, {and} \bibinfo{person}{Carey~S Ryan}.}
  \bibinfo{year}{2017}\natexlab{}.
\newblock \bibinfo{booktitle}{\emph{Data analysis: A model comparison approach
  to regression, ANOVA, and beyond}}.
\newblock \bibinfo{publisher}{Routledge}.
\newblock


\bibitem[Kapadia et~al\mbox{.}(2015)]%
        {kapadia2015evaluating}
\bibfield{author}{\bibinfo{person}{Mubbasir Kapadia}, \bibinfo{person}{Fabio
  Z{\"u}nd}, \bibinfo{person}{Jessica Falk}, \bibinfo{person}{Marcel Marti},
  \bibinfo{person}{Robert~W Sumner}, {and} \bibinfo{person}{Markus Gross}.}
  \bibinfo{year}{2015}\natexlab{}.
\newblock \showarticletitle{Evaluating the authoring complexity of interactive
  narratives with interactive behaviour trees}.
\newblock \bibinfo{journal}{\emph{Foundations of Digital Games}}
  (\bibinfo{year}{2015}).
\newblock


\bibitem[Karras et~al\mbox{.}(2017)]%
        {karras2017audio}
\bibfield{author}{\bibinfo{person}{Tero Karras}, \bibinfo{person}{Timo Aila},
  \bibinfo{person}{Samuli Laine}, \bibinfo{person}{Antti Herva}, {and}
  \bibinfo{person}{Jaakko Lehtinen}.} \bibinfo{year}{2017}\natexlab{}.
\newblock \showarticletitle{Audio-driven facial animation by joint end-to-end
  learning of pose and emotion}.
\newblock \bibinfo{journal}{\emph{ACM Transactions on Graphics (TOG)}}
  \bibinfo{volume}{36}, \bibinfo{number}{4} (\bibinfo{year}{2017}),
  \bibinfo{pages}{1--12}.
\newblock


\bibitem[Kopp et~al\mbox{.}(2006)]%
        {kopp2006towards}
\bibfield{author}{\bibinfo{person}{Stefan Kopp}, \bibinfo{person}{Brigitte
  Krenn}, \bibinfo{person}{Stacy Marsella}, \bibinfo{person}{Andrew~N
  Marshall}, \bibinfo{person}{Catherine Pelachaud}, \bibinfo{person}{Hannes
  Pirker}, \bibinfo{person}{Kristinn~R Th{\'o}risson}, {and}
  \bibinfo{person}{Hannes Vilhj{\'a}lmsson}.} \bibinfo{year}{2006}\natexlab{}.
\newblock \showarticletitle{Towards a common framework for multimodal
  generation: The behavior markup language}. In
  \bibinfo{booktitle}{\emph{International workshop on intelligent virtual
  agents}}. Springer, \bibinfo{pages}{205--217}.
\newblock


\bibitem[Lee and Marsella(2006)]%
        {lee2006nonverbal}
\bibfield{author}{\bibinfo{person}{Jina Lee} {and} \bibinfo{person}{Stacy
  Marsella}.} \bibinfo{year}{2006}\natexlab{}.
\newblock \showarticletitle{Nonverbal behavior generator for embodied
  conversational agents}. In \bibinfo{booktitle}{\emph{International Workshop
  on Intelligent Virtual Agents}}. Springer, \bibinfo{pages}{243--255}.
\newblock


\bibitem[Li et~al\mbox{.}(2021)]%
        {li2021controllable}
\bibfield{author}{\bibinfo{person}{Tao Li}, \bibinfo{person}{Shan Yang},
  \bibinfo{person}{Liumeng Xue}, {and} \bibinfo{person}{Lei Xie}.}
  \bibinfo{year}{2021}\natexlab{}.
\newblock \showarticletitle{Controllable emotion transfer for end-to-end speech
  synthesis}. In \bibinfo{booktitle}{\emph{2021 12th International Symposium on
  Chinese Spoken Language Processing (ISCSLP)}}. IEEE, \bibinfo{pages}{1--5}.
\newblock


\bibitem[Liu et~al\mbox{.}(2020)]%
        {liu2020building}
\bibfield{author}{\bibinfo{person}{Meng Liu}, \bibinfo{person}{Yaocong Duan},
  \bibinfo{person}{Robin~AA Ince}, \bibinfo{person}{Chaona Chen},
  \bibinfo{person}{Oliver~GB Garrod}, \bibinfo{person}{Philippe~G Schyns},
  {and} \bibinfo{person}{Rachael~E Jack}.} \bibinfo{year}{2020}\natexlab{}.
\newblock \showarticletitle{Building a generative space of facial expressions
  of emotions using psychological data-driven methods}. In
  \bibinfo{booktitle}{\emph{Proceedings of the 20th ACM International
  Conference on Intelligent Virtual Agents}}. \bibinfo{pages}{1--3}.
\newblock


\bibitem[Liu et~al\mbox{.}(2021)]%
        {liu2021speech}
\bibfield{author}{\bibinfo{person}{Yu Liu}, \bibinfo{person}{Gelareh
  Mohammadi}, \bibinfo{person}{Yang Song}, {and} \bibinfo{person}{Wafa Johal}.}
  \bibinfo{year}{2021}\natexlab{}.
\newblock \showarticletitle{Speech-based Gesture Generation for Robots and
  Embodied Agents: A Scoping Review}. In \bibinfo{booktitle}{\emph{Proceedings
  of the 9th International Conference on Human-Agent Interaction}}.
  \bibinfo{pages}{31--38}.
\newblock


\bibitem[Louwerse et~al\mbox{.}(2012)]%
        {louwerse2012behavior}
\bibfield{author}{\bibinfo{person}{Max~M Louwerse}, \bibinfo{person}{Rick
  Dale}, \bibinfo{person}{Ellen~G Bard}, {and} \bibinfo{person}{Patrick
  Jeuniaux}.} \bibinfo{year}{2012}\natexlab{}.
\newblock \showarticletitle{Behavior matching in multimodal communication is
  synchronized}.
\newblock \bibinfo{journal}{\emph{Cognitive science}} \bibinfo{volume}{36},
  \bibinfo{number}{8} (\bibinfo{year}{2012}), \bibinfo{pages}{1404--1426}.
\newblock


\bibitem[McCrae and John(1992)]%
        {mccrae1992introduction}
\bibfield{author}{\bibinfo{person}{Robert~R McCrae} {and}
  \bibinfo{person}{Oliver~P John}.} \bibinfo{year}{1992}\natexlab{}.
\newblock \showarticletitle{An introduction to the five-factor model and its
  applications}.
\newblock \bibinfo{journal}{\emph{Journal of personality}}
  \bibinfo{volume}{60}, \bibinfo{number}{2} (\bibinfo{year}{1992}),
  \bibinfo{pages}{175--215}.
\newblock


\bibitem[McDonnell et~al\mbox{.}(2008)]%
        {mcdonnell2008evaluating}
\bibfield{author}{\bibinfo{person}{Rachel McDonnell}, \bibinfo{person}{Sophie
  J{\"o}rg}, \bibinfo{person}{Joanna McHugh}, \bibinfo{person}{Fiona Newell},
  {and} \bibinfo{person}{Carol O'Sullivan}.} \bibinfo{year}{2008}\natexlab{}.
\newblock \showarticletitle{Evaluating the emotional content of human motions
  on real and virtual characters}. In \bibinfo{booktitle}{\emph{Proceedings of
  the 5th symposium on Applied perception in graphics and visualization}}.
  \bibinfo{pages}{67--74}.
\newblock


\bibitem[McNeill(1992)]%
        {mcneill1992hand}
\bibfield{author}{\bibinfo{person}{David McNeill}.}
  \bibinfo{year}{1992}\natexlab{}.
\newblock \showarticletitle{Hand and Mind: What Gestures Reveal About Thought.}
\newblock  (\bibinfo{year}{1992}).
\newblock


\bibitem[Nagy et~al\mbox{.}(2021)]%
        {nagy2021framework}
\bibfield{author}{\bibinfo{person}{Rajmund Nagy}, \bibinfo{person}{Taras
  Kucherenko}, \bibinfo{person}{Birger Moell}, \bibinfo{person}{Andr{\'e}
  Pereira}, \bibinfo{person}{Hedvig Kjellstr{\"o}m}, {and}
  \bibinfo{person}{Ulysses Bernardet}.} \bibinfo{year}{2021}\natexlab{}.
\newblock \showarticletitle{A framework for integrating gesture generation
  models into interactive conversational agents}.
\newblock \bibinfo{journal}{\emph{arXiv preprint arXiv:2102.12302}}
  (\bibinfo{year}{2021}).
\newblock


\bibitem[Nvidia(2021)]%
        {omniverse2021}
\bibfield{author}{\bibinfo{person}{Nvidia}.} \bibinfo{year}{2021}\natexlab{}.
\newblock \bibinfo{title}{Omniverse Audio2Face}.
\newblock
\newblock


\bibitem[Potdevin et~al\mbox{.}(2018)]%
        {potdevin2018virtual}
\bibfield{author}{\bibinfo{person}{Delphine Potdevin},
  \bibinfo{person}{C{\'e}line Clavel}, {and} \bibinfo{person}{Nicolas
  Sabouret}.} \bibinfo{year}{2018}\natexlab{}.
\newblock \showarticletitle{Virtual Intimacy, this little something between us:
  a study about Human perception of intimate behaviors in Embodied
  Conversational Agents}. In \bibinfo{booktitle}{\emph{Proceedings of the 18th
  international conference on intelligent virtual agents}}.
  \bibinfo{pages}{165--172}.
\newblock


\bibitem[Qualtrics(2021)]%
        {qualtrics}
\bibfield{author}{\bibinfo{person}{Qualtrics}.}
  \bibinfo{year}{2021}\natexlab{}.
\newblock \bibinfo{booktitle}{\emph{Qualtrics}}.
\newblock Qualtrics, Provo, Utah, USA.
\newblock
\urldef\tempurl%
\url{http://www.qualtrics.com}
\showURL{%
\tempurl}


\bibitem[Randhavane et~al\mbox{.}(2019)]%
        {randhavane2019eva}
\bibfield{author}{\bibinfo{person}{Tanmay Randhavane}, \bibinfo{person}{Aniket
  Bera}, \bibinfo{person}{Kyra Kapsaskis}, \bibinfo{person}{Rahul Sheth},
  \bibinfo{person}{Kurt Gray}, {and} \bibinfo{person}{Dinesh Manocha}.}
  \bibinfo{year}{2019}\natexlab{}.
\newblock \showarticletitle{Eva: Generating emotional behavior of virtual
  agents using expressive features of gait and gaze}. In
  \bibinfo{booktitle}{\emph{ACM symposium on applied perception 2019}}.
  \bibinfo{pages}{1--10}.
\newblock


\bibitem[Ravenet et~al\mbox{.}(2018)]%
        {ravenet2018automating}
\bibfield{author}{\bibinfo{person}{Brian Ravenet}, \bibinfo{person}{Catherine
  Pelachaud}, \bibinfo{person}{Chlo{\'e} Clavel}, {and} \bibinfo{person}{Stacy
  Marsella}.} \bibinfo{year}{2018}\natexlab{}.
\newblock \showarticletitle{Automating the production of communicative gestures
  in embodied characters}.
\newblock \bibinfo{journal}{\emph{Frontiers in psychology}}
  \bibinfo{volume}{9} (\bibinfo{year}{2018}), \bibinfo{pages}{1144}.
\newblock


\bibitem[Russell(1980)]%
        {russell1980circumplex}
\bibfield{author}{\bibinfo{person}{James~A Russell}.}
  \bibinfo{year}{1980}\natexlab{}.
\newblock \showarticletitle{A circumplex model of affect.}
\newblock \bibinfo{journal}{\emph{Journal of personality and social
  psychology}} \bibinfo{volume}{39}, \bibinfo{number}{6}
  (\bibinfo{year}{1980}), \bibinfo{pages}{1161}.
\newblock


\bibitem[Sajjadi et~al\mbox{.}(2019)]%
        {sajjadi2019personality}
\bibfield{author}{\bibinfo{person}{Pejman Sajjadi}, \bibinfo{person}{Laura
  Hoffmann}, \bibinfo{person}{Philipp Cimiano}, {and} \bibinfo{person}{Stefan
  Kopp}.} \bibinfo{year}{2019}\natexlab{}.
\newblock \showarticletitle{A personality-based emotional model for embodied
  conversational agents: Effects on perceived social presence and game
  experience of users}.
\newblock \bibinfo{journal}{\emph{Entertainment Computing}}
  \bibinfo{volume}{32} (\bibinfo{year}{2019}), \bibinfo{pages}{100313}.
\newblock


\bibitem[Schr{\"o}der(2001)]%
        {schroder2001emotional}
\bibfield{author}{\bibinfo{person}{Marc Schr{\"o}der}.}
  \bibinfo{year}{2001}\natexlab{}.
\newblock \showarticletitle{Emotional speech synthesis: A review}. In
  \bibinfo{booktitle}{\emph{Seventh European Conference on Speech Communication
  and Technology}}. Citeseer.
\newblock


\bibitem[Shoulson et~al\mbox{.}(2013)]%
        {shoulson2013adapt}
\bibfield{author}{\bibinfo{person}{Alexander Shoulson}, \bibinfo{person}{Nathan
  Marshak}, \bibinfo{person}{Mubbasir Kapadia}, {and} \bibinfo{person}{Norman~I
  Badler}.} \bibinfo{year}{2013}\natexlab{}.
\newblock \showarticletitle{Adapt: the agent developmentand prototyping
  testbed}.
\newblock \bibinfo{journal}{\emph{IEEE Transactions on Visualization and
  Computer Graphics}} \bibinfo{volume}{20}, \bibinfo{number}{7}
  (\bibinfo{year}{2013}), \bibinfo{pages}{1035--1047}.
\newblock


\bibitem[Shvo et~al\mbox{.}(2019)]%
        {shvo2019interdependent}
\bibfield{author}{\bibinfo{person}{Maayan Shvo}, \bibinfo{person}{Jakob
  Buhmann}, {and} \bibinfo{person}{Mubbasir Kapadia}.}
  \bibinfo{year}{2019}\natexlab{}.
\newblock \showarticletitle{An interdependent model of personality, motivation,
  emotion, and mood for intelligent virtual agents}. In
  \bibinfo{booktitle}{\emph{Proceedings of the 19th ACM International
  Conference on Intelligent Virtual Agents}}. \bibinfo{pages}{65--72}.
\newblock


\bibitem[Sohn et~al\mbox{.}(2018)]%
        {sohn2018emotionally}
\bibfield{author}{\bibinfo{person}{Samuel~S Sohn}, \bibinfo{person}{Xun Zhang},
  \bibinfo{person}{Fernando Geraci}, {and} \bibinfo{person}{Mubbasir Kapadia}.}
  \bibinfo{year}{2018}\natexlab{}.
\newblock \showarticletitle{An emotionally aware embodied conversational
  agent}. In \bibinfo{booktitle}{\emph{Proceedings of the 17th International
  Conference on Autonomous Agents and MultiAgent Systems}}.
  \bibinfo{pages}{2250--2252}.
\newblock


\bibitem[Sonlu et~al\mbox{.}(2021)]%
        {sonlu2021conversational}
\bibfield{author}{\bibinfo{person}{Sinan Sonlu}, \bibinfo{person}{U{\u{g}}ur
  G{\"u}d{\"u}kbay}, {and} \bibinfo{person}{Funda Durupinar}.}
  \bibinfo{year}{2021}\natexlab{}.
\newblock \showarticletitle{A conversational agent framework with multi-modal
  personality expression}.
\newblock \bibinfo{journal}{\emph{ACM Transactions on Graphics (TOG)}}
  \bibinfo{volume}{40}, \bibinfo{number}{1} (\bibinfo{year}{2021}),
  \bibinfo{pages}{1--16}.
\newblock


\bibitem[Wagner et~al\mbox{.}(2014)]%
        {wagner2014gesture}
\bibfield{author}{\bibinfo{person}{Petra Wagner}, \bibinfo{person}{Zofia
  Malisz}, {and} \bibinfo{person}{Stefan Kopp}.}
  \bibinfo{year}{2014}\natexlab{}.
\newblock \bibinfo{title}{Gesture and speech in interaction: An overview}.
\newblock , \bibinfo{numpages}{209--232}~pages.
\newblock


\bibitem[Welch(1947)]%
        {welch1947generalization}
\bibfield{author}{\bibinfo{person}{Bernard~L Welch}.}
  \bibinfo{year}{1947}\natexlab{}.
\newblock \showarticletitle{The generalization of ‘STUDENT'S’problem when
  several different population varlances are involved}.
\newblock \bibinfo{journal}{\emph{Biometrika}} \bibinfo{volume}{34},
  \bibinfo{number}{1-2} (\bibinfo{year}{1947}), \bibinfo{pages}{28--35}.
\newblock


\bibitem[Wolfert et~al\mbox{.}(2022)]%
        {wolfert2022review}
\bibfield{author}{\bibinfo{person}{Pieter Wolfert}, \bibinfo{person}{Nicole
  Robinson}, {and} \bibinfo{person}{Tony Belpaeme}.}
  \bibinfo{year}{2022}\natexlab{}.
\newblock \showarticletitle{A review of evaluation practices of gesture
  generation in embodied conversational agents}.
\newblock \bibinfo{journal}{\emph{IEEE Transactions on Human-Machine Systems}}
  (\bibinfo{year}{2022}).
\newblock


\bibitem[Yal{\c{c}}{\i}n(2020)]%
        {yalccin2020empathy}
\bibfield{author}{\bibinfo{person}{{\"O}zge~Nilay Yal{\c{c}}{\i}n}.}
  \bibinfo{year}{2020}\natexlab{}.
\newblock \showarticletitle{Empathy framework for embodied conversational
  agents}.
\newblock \bibinfo{journal}{\emph{Cognitive Systems Research}}
  \bibinfo{volume}{59} (\bibinfo{year}{2020}), \bibinfo{pages}{123--132}.
\newblock


\bibitem[Yoon et~al\mbox{.}(2020)]%
        {yoon2020speech}
\bibfield{author}{\bibinfo{person}{Youngwoo Yoon}, \bibinfo{person}{Bok Cha},
  \bibinfo{person}{Joo-Haeng Lee}, \bibinfo{person}{Minsu Jang},
  \bibinfo{person}{Jaeyeon Lee}, \bibinfo{person}{Jaehong Kim}, {and}
  \bibinfo{person}{Geehyuk Lee}.} \bibinfo{year}{2020}\natexlab{}.
\newblock \showarticletitle{Speech gesture generation from the trimodal context
  of text, audio, and speaker identity}.
\newblock \bibinfo{journal}{\emph{ACM Transactions on Graphics (TOG)}}
  \bibinfo{volume}{39}, \bibinfo{number}{6} (\bibinfo{year}{2020}),
  \bibinfo{pages}{1--16}.
\newblock


\bibitem[Yoon et~al\mbox{.}(2019)]%
        {yoonICRA19}
\bibfield{author}{\bibinfo{person}{Youngwoo Yoon}, \bibinfo{person}{Woo-Ri Ko},
  \bibinfo{person}{Minsu Jang}, \bibinfo{person}{Jaeyeon Lee},
  \bibinfo{person}{Jaehong Kim}, {and} \bibinfo{person}{Geehyuk Lee}.}
  \bibinfo{year}{2019}\natexlab{}.
\newblock \showarticletitle{Robots Learn Social Skills: End-to-End Learning of
  Co-Speech Gesture Generation for Humanoid Robots}. In
  \bibinfo{booktitle}{\emph{Proc. of The International Conference in Robotics
  and Automation (ICRA)}}.
\newblock


\bibitem[Zhou et~al\mbox{.}(2018)]%
        {Zhou2018EmotionalCM}
\bibfield{author}{\bibinfo{person}{Hao Zhou}, \bibinfo{person}{Minlie Huang},
  \bibinfo{person}{Tianyang Zhang}, \bibinfo{person}{Xiaoyan Zhu}, {and}
  \bibinfo{person}{Bing-Qian Liu}.} \bibinfo{year}{2018}\natexlab{}.
\newblock \showarticletitle{Emotional Chatting Machine: Emotional Conversation
  Generation with Internal and External Memory}. In
  \bibinfo{booktitle}{\emph{AAAI}}.
\newblock


\end{thebibliography}


\end{document}


\title[The Importance of Multimodal Emotion Conditioning and Affect Consistency for Embodied Conversational Agents]{Appendix to: The Importance of Multimodal Emotion Conditioning and Affect Consistency for Embodied Conversational Agents}







\maketitle



\appendix
\section{Implementation Details of the Framework} \label{app:framework_detail}
\begin{figure}[h]
  \centering
  \includegraphics[width=\linewidth]{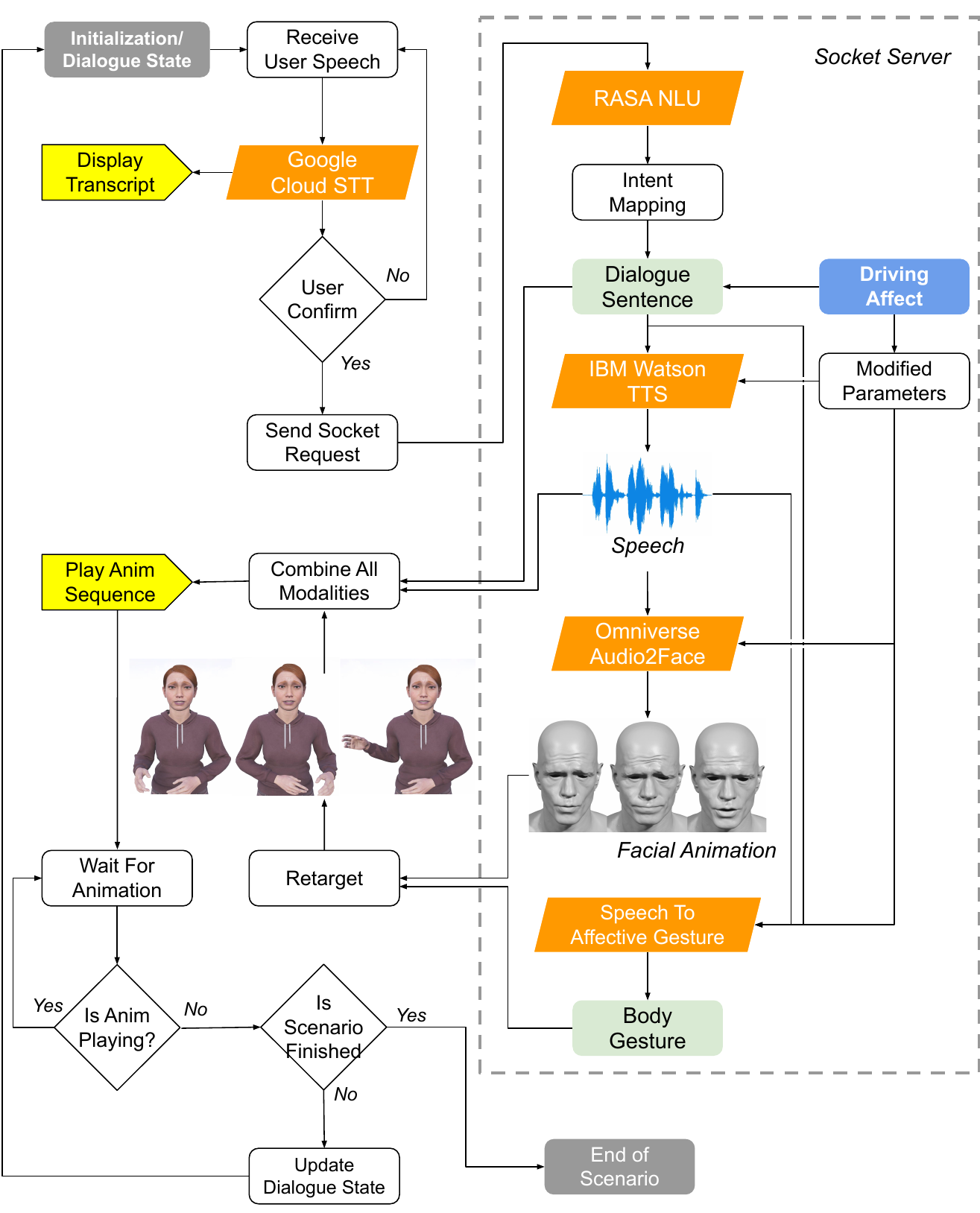}
  \caption{The flowchart of our framework for dialogue handling and multimodal behavior generation. The handler starts by receiving input audio from the user (upper left). After the transcription is confirmed, a socket request is sent to the server. Then the client receives the multimodal behaviors. The animation is played afterwards and the dialogue handler repeats until the end of scenario. Note that the dialogue manager generates the text response according to the driving affect in the socket server. The dependencies of the other modalities are as follows. The emotional audio is synthesized from the text, the facial animation is generated from audio, and the body movement is from both text and audio.}
  \label{fig:framework}
\end{figure}

The implementation details of our framework are shown in Figure \ref{fig:framework}. For the perception module, we use the Google Cloud Speech-To-Text \cite{gcloudstt} followed by a trained RASA NLU module \cite{bocklisch2017rasa} for user intent extraction. In our dialogue manager, we map the intent to the corresponding text response, given the driving affect. The text and the modified affective features are then passed to IBM Text-To-Speech Engine for emotional speech synthesis. 
The synthetic audio, along with the modified affective parameters, are then used to generate expressive facial animation with Ominverse Audio2Face \cite{omniverse2021}. The text, audio, and the driving affect, are used to generate affective gestures. 

We build our framework based on the high-fidelity parameterized digital character, Metahuman \cite{metahuman2021}. 
The management of the character's perception and responses is implemented in Unreal Engine 4 \cite{ue4}.
A socket server-client connection is constructed, where the blueprint client sends the perceived audio to our socket server and the multimodal response is received for animation display.

\section{Scenario Flowchart} \label{app:scenarios}
\begin{figure*}[!h] 
  \centering
  \includegraphics[width=\linewidth]{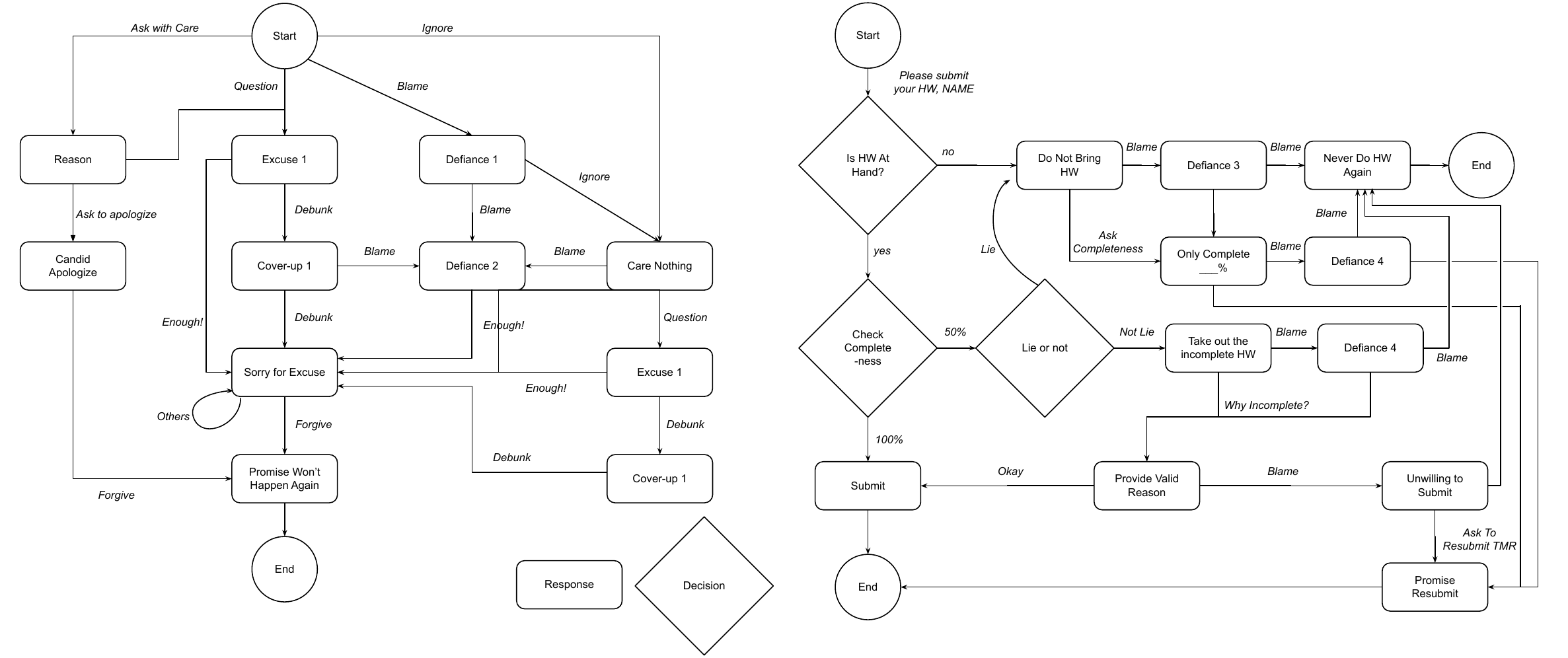}
  \caption{Our designed late (left) and homework (right) scenarios. Dialogue states are represented as rectangles and the transitions between the states are represented as arrows. The transitions of the dialogues are dependent on the intent extracted from user's speech. At each dialogue state, our framework can generate the affective multimodal responses.}
  \label{fig:scenarios}
\end{figure*}

Figure \ref{fig:scenarios} shows our late and homework scenarios. 
The transition of dialogue states are determined by the extracted intent from user's speech. For each dialogue state, the 6 parallel affective sentences are prepared. Based on the driving affect, the emotional voice, face, and body animations will be synthesized accordingly. It's also noted that the flowchart of our homework scenario is parameterized by three variables, specifying whether HW is at hand, the HW completeness, and whether or not to lie. The transitions of some dialogue states are dependent on the variables, which makes the scenario more complex.

\begin{figure*}[!h]
  \centering
  \begin{subfigure}[b]{0.42\linewidth}
    \centering
    \includegraphics[width=\linewidth]{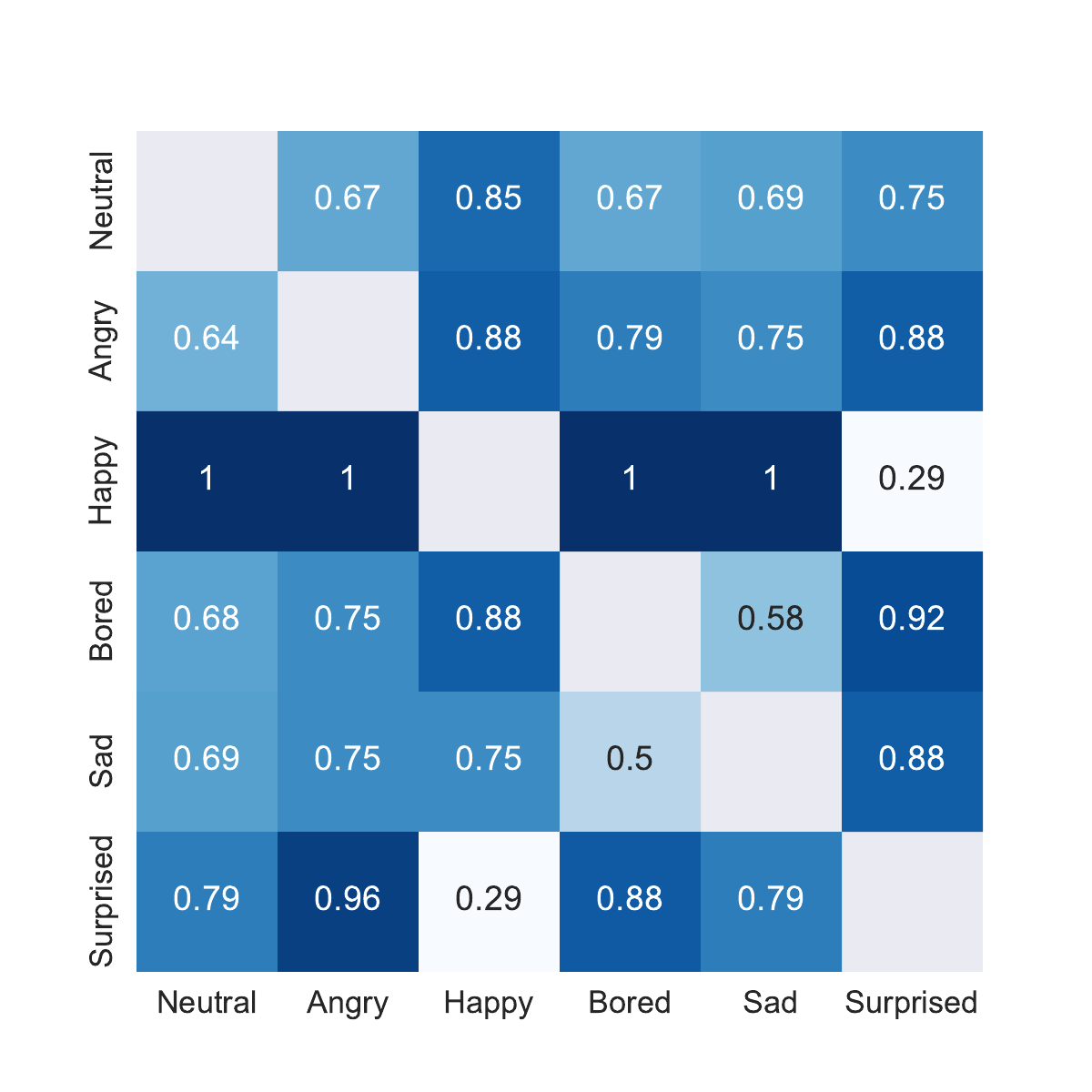}
    \caption{Face modality.}
  \label{fig:affect_cm_face}
  \end{subfigure}
  \hfill
    \begin{subfigure}[b]{0.42\linewidth}
    \centering
    \includegraphics[width=\linewidth]{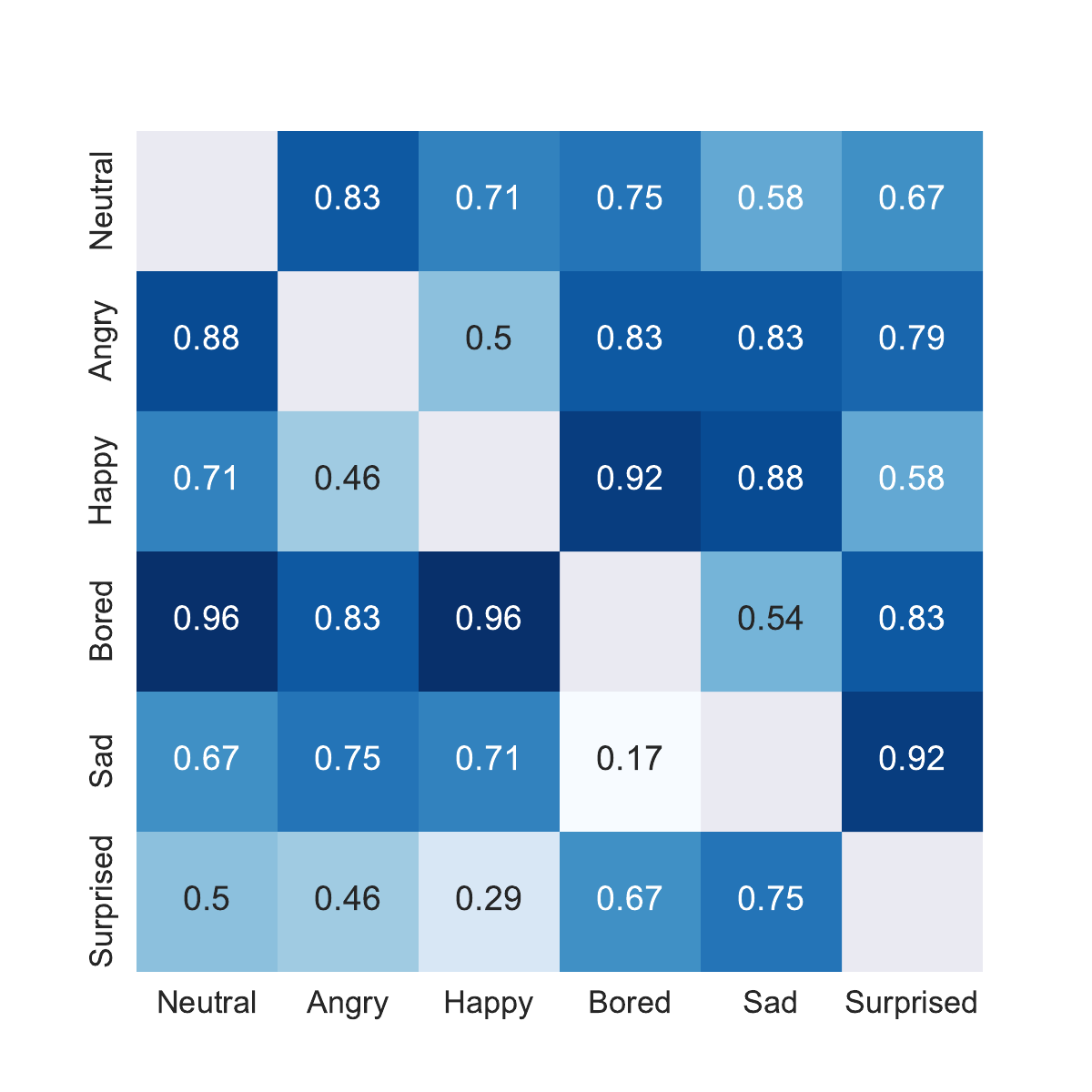}
    \caption{Body modality.}
    \label{fig:affect_cm_body}
  \end{subfigure}
  \caption{Average match for pairwise comparison for affects in the user study. The slot values represent the average match of a pair of affects at the X and Y axes when the affect at the Y axis is queried.} \label{fig:affect_cm}
\end{figure*}


\section{Additional Results for Preliminary User Study} \label{app:additional_pre_userstudy}
We further calculate the average match (match + 0.5 * equal) for all paired affects to see which pairs of affective behaviors are perceptually correlated. Figure \ref{fig:affect_cm} shows the two confusion matrices for the face and body modality. The slot values indicate the average match when the two affective bahaviors, one from X-axis and the other from Y-axis, are presented and the affect at the Y-axis is queried. The higher the value is, the more confident the participants are in saying the affect at the Y-axis is perceived more like the affect at the Y-axis than the affect at the X-axis. The matrices are not completely symmetric along the diagonal because the users can select asymmetric options for both affect queries.

For the face modality (Figure \ref{fig:affect_cm_face}), happiness and surprise are hard for the users to tell apart. The average match is only 29\%. When the happy faces are put together with the other affective faces, users can always identify them. Bored and sad facial animations are also perceptually similar. Apart from that, our selection of affective facial parameters leads to distinguishable facial behaviors for most pairs.

For the body modality (Figure \ref{fig:affect_cm_body}), happiness and anger are hard to tell apart, with their average matches below 50\%. Sad and bored gestures can also be confusing to our participants. The surprised body motions seem hard to generate as they can only be correctly identified when paired with bored or sad gestures. From the confusion matrix, we can roughly cluster angry and happy gestures as one group whereas bored and sad as the other group. The affects of the same cluster are distinguishable from other different affects. For example, bored and sad gestures can be both identified when paired with neutral, angry, happy, and surprised motions.

\bibliographystyle{ACM-Reference-Format}
\bibliography{reference}